\documentclass[%
 reprint,
superscriptaddress,
 amsmath,amssymb,
 aps,
pra,
]{revtex4-2}

\usepackage{graphicx}
\usepackage{dcolumn}
\usepackage{bm}
\usepackage[dvipsnames]{xcolor}
\usepackage{appendix}

\newcommand{\ket}[1]{\vert #1 \rangle}
\newcommand{\bra}[1]{\langle #1 \vert}

\begin{document}

\preprint{APS/123-QED}

\title{Tutorial: Optical quantum metrology}

\author{Marco Barbieri}
\affiliation{
 Dipartimento di Scienze, Universit\`a degli Studi Roma Tre, Via della Vasca Navale, 84, 00146 Rome, Italy
}
\affiliation{
 Istituto Nazionale di Ottica, CNR, Largo Enrico Fermi 6, 50125 Florence, Italy
}


\begin{abstract}
The purpose of quantum technologies is to explore how quantum effects can improve on existing solutions for the treatment of information. Quantum photonics sensing holds great promises for reaching a more efficient trade-off between invasivity and quality of the measurement, when compared with the potential of classical means. This tutorial is dedicated to presenting how this advantage is brought about by nonclassical light, examining the basic principles of parameter estimation and reviewing the state of the art.
\end{abstract}

\maketitle


\section{Introduction}

Measurements are physical processes. As such, their power is constrained by requirements and limitations, no more and no less than the same phenomena they observe. There exist a long list of occurrences: the internal resistance of an ammeter is meant to be small compared to the load, while that of a voltmeter should be large, the heat capacity of a thermometer should be small, and so on and so forth. In addition, all measurement devices will introduce some form of noise: for instance, any resistance will present fluctuations of the current, due to the thermal motion of its electrons. 

All these facts are often recited out concluding that the resulting disturbance can be made, in principle, arbitrarily small. However, when trying and applying this principle to real cases, other considerations may come into play. While the signal-to-noise ratio in an optical absorption measurement could be made large at will by ramping up the intensity on the sample, in practice this may affect the specimen or saturate the amplifier. This is in fact a common instance in optical measurements: they typically improve with the intensity, however they may alter  the sample by delivering energy to it. There may occur, then, a trade off between the quality of the measurement, and its invasivity. Thus, the promised arbitrariness in noise rejection must actually come to terms with contrasting considerations. 

The most fundamental limitations on measurements are to be sought in the most fundamental theory of matter: quantum mechanics. Once all instrumental causes of error and disturbance are removed, we are left with the fluctuations inherent to quantum states as the source of variability in our measurements. Understanding what these restrictions are and what opportunities open up is the aim of quantum metrology~\cite{Giovannetti:2011yq}. Controlling the wavefunction of the object employed as the probe does not make the trade off disappear, but it provides the means for a more satisfactory compromise, whenever the simplest option - more power - can not be adopted. In this tutorial, we wish to discuss the essentials of quantum metrology in its applications to photonics~\cite{Pirandola:2018qd,doi:10.1116/5.0007577}. Along with solid-state~\cite{RevModPhys.89.035002} and atomic systems~\cite{RevModPhys.90.035005}, this represent one of the most investigated platforms, due to its adaptability and the perspective of leveraging on solutions for current photonics sensors. This tutorial is intended for those searching an introduction to the topic and a discussion of its methods and concepts, coming from different backgrounds. It thus resembles in its purposes to the Tube map: it guides the traveller though the main stops, but it should not be taken as a detailed outline. And it does not aim at coming any close to the Knowledge.

\section{Fundamentals}

\subsection{The conceptual framework}

The uncertainty relation~\cite{Heisenberg49}, writing in its canonical form
\begin{equation}
\label{unrel}
    \Delta x\, \Delta p \geq \frac{\hbar}{2}
\end{equation}
for a pair of conjugate observables $\hat x$ and $\hat p$, is one of the characteristic traits of quantum mechanics, and one of the most abused tricks in the the hands of popularizers when it comes to presenting the weirdness of the quantum world to the large public. Although the attitude of the specialist is expected to be more detached and analytical, many would confess this concept retains a certain fascination, even after years of  practice in the field.

\begin{figure}[h]
    \centering
    \includegraphics[width=\columnwidth]{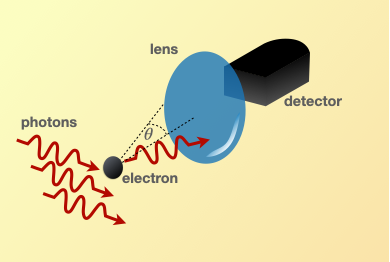}
    \caption{Heisenberg's {\it gedankenexperiment}. It summarises all aspects to be used in analysing the measurement process.}
    \label{fig:Heisenberg}
\end{figure}

Albeit a simple formula, the relation \eqref{unrel} is prone to different interpretations. At a very abstract level, it is a consequence of adopting square-integrable functions as physical wavefunctions, whose Fourier transform is then bound to satisfy this constraint. In classical physics, this statement is translated as the observation that an oscillation can be localised in a region of size $\Delta x$ by superposing plane waves spanning over the range of wavevectors $\Delta k = \Delta p/\hbar$. Thus, translating this to quantum states, we could conclude that \eqref{unrel} sets limits to our capabilities in \emph{state preparation}. 

It is  remarkable that Heisenberg deemed it instructive to provide an interpretation from the point of view of the measurement, instead. His celebrated thought experiment of a single-photon microscope detecting a single electron is sketched in Fig.~\ref{fig:Heisenberg}. Imagine we are interested in finding the position of an electron; for our task, we collect a scattered photon through a microscope lens. The complete arrangement then consists of the system to be measured - the electron, a probe by which we carry out our observation - the photon, and, finally, a measurement setup to finalise our observation.

The interaction process is  a Compton scattering event, by which the photon deviates from its original path; the position of the electron can be inferred by measuring the photon in its new direction.
At best, the electron can be localised within the optical resolution of the microscope: $\Delta x = \lambda/\sin\theta$, with $\theta$ the maximal collection angle. This can be set by choosing the wavelength of the photon and the numerical aperture of the lens. It should be remarked that the position of the electron is accessed by means of a measurement on a different system, the photon, as it enters as a parameter in the detection probability distribution of the latter. On the other hand, scattering leads to the electron receiving a recoil, modifying its momentum within a spread $\Delta p = 4\pi \hbar\sin\theta/\lambda$: The interaction has affected the wavefunction of the electron.
The product $\Delta x\,\Delta p$ is thus close to the rigorous result \eqref{unrel}. 

We can now ponder on how to sit on Heisenberg's shoulders, and look for a framework that could apply in broader scenarios. This requires to analyse Heisenberg's thought experiment under a more abstract point of view. First, we consider how the electron should be illuminated, bearing in mind that the wavelength dictates the optical resolution. Also, the incoming direction of the photon should be set properly, in such a way not to cause stray light: this would not carry any signature of the electron's position, and would end up decreasing the signal-to-noise ratio of our measurement. These considerations highlight the first critical step as the {\it preparation} of the probe, {\it i.e.} the physical system used for monitoring.

Next, we consider that, if we wish to obtain the position of the electron from a measurement on the photon, we  have to rely on our  knowledge that Compton scattering is occurring. We have thus learned the physical mechanism behind the measurement: we can make exact predictions, based on the knowledge of the {\it interaction}, be it in the form of either a unitary or a dissipative process. This, in turn, is characterised by one or more parameters which are the quantities we want to estimate with the best possible precision.

Finally, the physical size of the lens curtails our ability of localising the electron.
In general terms, after the interaction, the probe is delivered to a measuring apparatus designed to access one observable of the probe. There exist physical limitations on the quality of this {\it measurement} dictating what we can actually learn about the parameters.   
The complete process that eventually delivers the estimate of the electron position thus consists of the triad preparation-interaction-measurement.

\begin{figure}
    \centering
    \includegraphics[width=\columnwidth]{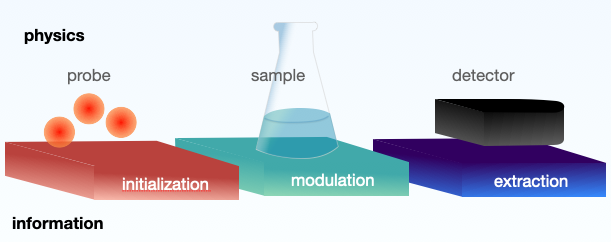}
    \caption{Conceptual scheme of the estimation procedure, highlighting the main elements physical implementation, and the corresponding steps in information processing}
    \label{fig:triad}
\end{figure}

We are now prepared to take one step beyond in abstraction and read the triad above as the physical implementation of an information exchange~\cite{T_th_2014}. This starts with the {\it initialisation} of the probe to a blank state on which information about the sought parameters will be written. The interaction effects a {\it modulation}, which depends on the values of the parameters following a known law. This eventually allows to proceed with information {\it extraction}, with a varying degree of effectiveness. We can enlarge this construction to include cases in which the parameters pertain directly to the initial state, as, for instance, the level of entanglement or the purity~\cite{PhysRevLett.104.100501,Virzi:2019qy}, by including the interaction stage as a part of the state preparation. Beneath this all, quantum mechanics puts limits to what is physically possible or, equivalently, on the information exchange. These lines of reasoning are summarised in Fig.~\ref{fig:triad}. 

For a given parameter, the choice of the probe state and the measurement is what we call a {\it strategy}. The aim is to optimise the amount of information this carries on the value of the parameters. Intuition guides us to conclude that the most informative measurements are the ones resulting in the lowest uncertainty. Any comparison, however, only makes sense if the resources invested are kept fixed. This is a key concept, but translating it to a rigorous definition is perhaps elusive. It is maybe best to adhere to an all-encompassing description, calling a resource anything that is useful to extract information. These include the physical constituents of the probe, the number of times the system is accessed in a single experimental run, and the total number of runs. 

\subsection{Measuring light}
Before dwelling on  more sophisticated aspects of quantum measurements, we take some time to discuss the more profane matter of how one actually measures properties of light in the laboratory. First, one should acknowledge that quantum optics often requires a certain mental agility in passing from the first-quantisation to the second-quantisation picture. This is not solely a matter of making proper calculations, but, crucially, to understand what is actually measured. Different kinds of detectors rely on different physical mechanisms for the measurement, and it is important to delineate exactly how these are related to the properties of quantum light.
A short summary on the manipulation of quantum states of light is presented in Appendix A.

\subsubsection{Photon counting}

The simplest detection scheme is linear intensity detection by means of a photodiode. This device converts the luminous flow to a proportional electronic current, up to loss from reflections on the elements of the detectors (e.g. protective windows or the active area itself) and from the elementary electron scattering mechanism. When operated on a classical intense beam, the current will be affected, at best, only by the shot noise~\cite{Schottky}, {i.e.} random Poisson fluctuations of the photon number showing up in the current~\cite{Bachor}. More realistically, there will also be an electronic noise component, associated to their thermal excitation, and extra fluctuations due to unwanted modulations of the beam intensity. These often prevent from operating this measurement scheme at the ultimate precision limit, hence direct detection is hardly ever used in quantum metrology by itself. 

\begin{figure}
    \centering
    \includegraphics[width=\columnwidth]{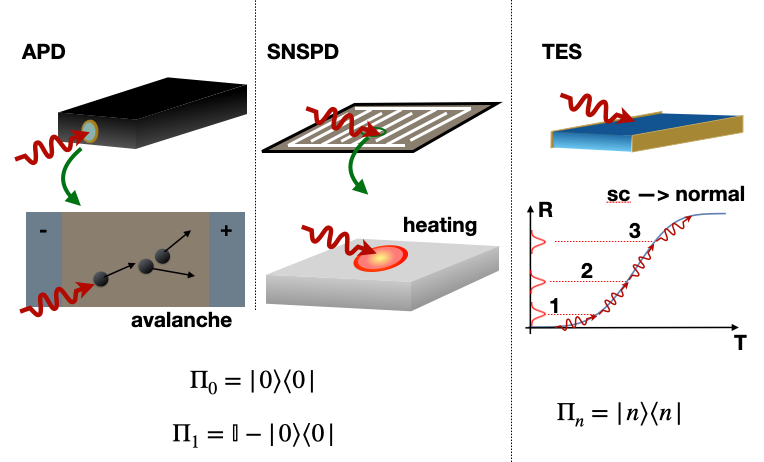}
    \caption{Photon counting detectors. The Avalanche PhotoDiode (APD) relies on an avalanche effect to form the signal. The Superconducting Nanowire Single Photon Detector (SNSPD) detects a photon from the resulting heating which disrupts superconductivity. Both detectors can be described by the on/off measurement operators $\Pi_0$ and $\Pi_1$ in the ideal limit of zero noise and unit efficiency. The Transition Edge Detector(TES) relies on the breaking of superconductivity as well, but the resulting resistance R is a function of its effective temperature T, thus of the total impinging energy. It thus realises a photon number discriminator with measurement operators $\Pi_n$ in the same ideal limit.}
    \label{fig:detectors}
\end{figure}

Linear detectors can not give access to single photons: the corresponding current would be inappreciable due to the electronic noise. This is why avalanche photodiodes (APDs) are employed (Fig.~\ref{fig:detectors}): in these detectors, the excitation of even a single electron is able to empty the active area of all free charges by means of successive collisions~\cite{Saleh:1084451}. This avalanche effect is achieved by polarising the detector junction in reverse bias. The signal is sufficiently high that it can be detected, however it carries no information on the actual photon number impinging on the detector: as a terse summary, even a single photon  suffices to saturate the current from the detector. This scheme thus only provides an on/off event - often called a `click'. The intrinsic quantum efficiency of these devices is around 60\% in the visible (500nm - 800nm) for Si detectors, polluted by dark counts due to thermal activation of the order of 1000 events/s. For longer wavelengths, different semiconductors are more indicated, reaching efficiencies of the order of 20\%~\cite{Eisaman}. These detectors are not mode sensitive: they would respond with clicks to whichever mode reaches them, compatibly with their spectral characteristics. In order to restrict their response to the few modes in which one is typically interested, being them spatial or spectral, filters are employed, at the cost of bringing the efficiency further down. 

More information on the photon number is collected by using multiple `click detectors', going through the effort of dividing the initial beam over multiple spatial or temporal bins~\cite{PhysRevA.68.043814, doi:10.1080/09500340408235288}. Care must be taken when handling the outputs of such detectors~\cite{PhysRevLett.109.093601}: while counting abilities are improved, the convergence of the click distribution to the actual photon number distribution is slow with the total number of bins.  Superconducting nanowire detectors offer superior efficiencies~\cite{nanowirereview}: their working principle is based on the fact that the energy of a single photon is sufficient to disrupt superconductivity in one section of the wire (Fig.\ref{fig:detectors}). If light illuminates the wire uniformly, one effectively realises a multiplexed click detector by an effective spatial binning~\cite{Divochiy:2008ij,Zhu:2020mb}, but time multiplexing can also be employed~\cite{Natarajan:13}. This offers a more compact solution than simple multiplexing, with an intrinsic quantum efficiency of the order of 95\%~\cite{Marsili:2013it,Reddy:20}. 

Genuine photon number resolution can be obtained based on superconducting transition edge sensors~\cite{Lita:08}: these detectors are bolometers, kept close to the transition temperature and whose signal is related to a disturbance to the superconductivity, as in nanowires (Fig.\ref{fig:detectors}). If the effect of the light is not strong enough to drive the detector to its normal state, a resistance is measured, proportional to the impinging energy; at fixed wavelength, this amounts to counting the number of photons. Intrinsic quantum efficiencies exceed 95\%, with the ability of distinguishing 10 to 20 photons, as a typical value~\cite{Humphreys_2015}; with higher energies, the normal state is fully reached, and saturation occurs. As for other counting systems, a signal is obtained regardless the mode of the incoming photons, hence the need of  filtering the desired modes.

\subsubsection{Coherent detection}

Photon counting addresses particle-like aspects of light, or, more properly, of its energy exchange. Alongside with these, there exist wave-like properties, which are typically accessed by means of interference. Taking inspiration from classical signal processing, the electric field of an optical mode can be decomposed in two components called quadratures: $\hat x$, which is in phase with a given local oscillator, and $\hat p$ which has a $\pi/2$-shift. In terms of quantum operators, these two are related to the 
creation-annihilation operators $\hat a$ and $\hat a^\dag$   as~\cite{Bachor,Loudon:105699,Serafozzi} $\hat x =\sqrt{N_0}(\hat a^\dag+\hat a)$ and $\hat p =i\sqrt{N_0}(\hat a^\dag-\hat a)$, and satisfy the commutation relation $[\hat x,\hat p]=2iN_0$. These are the canonical variables describing an electromagnetic mode as a quantum harmonic oscillator. 

For a coherent state $\vert \alpha \rangle$, we find $\langle \alpha\vert \hat x\vert \alpha \rangle=2\sqrt{N_0}\text{Re}[\alpha]$, and
$\langle \alpha\vert \hat p\vert \alpha \rangle=2\sqrt{N_0}\text{Im}[\alpha]$, hence reinforcing the view of $\hat x$ and $\hat p$ as the in-phase and in-quadrature field components respectively. In these states, the variances are given by $\Delta^2 x =\Delta^2 p = N_0$, independent on the amplitude $\alpha$. Therefore, these are the characteristic fluctuations associated to the vacuum ($\alpha=0$)~\cite{walls1995quantum,Loudon:105699,Bachor}. The constant $N_0$ determines the convention for the units, and many are used in the literature: the most common ones set $N_0=1/2$ (the relations between above become symmetric), $N_0=1$ ( the quadratures are normalised to the fluctuations in the vacuum), or $N_0=1/4$ (the conversion factor between expectation values and $\alpha$ is one). In the following, we will use $N_0=1/2$. 

\begin{figure}
    \centering
    \includegraphics[width=\columnwidth]{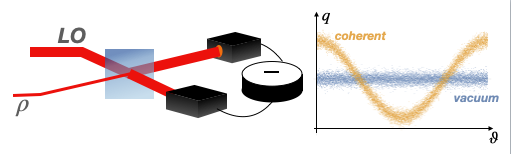}
    \caption{Homodyne detection makes it possible to measure the distribution of the generalised quadratures $\hat q$ of the state $\rho$, by means of interference with a local oscillator LO, whose phase $\vartheta$ can be varied. The vacuum state $\rho=\vert 0\rangle\langle 0\vert$ shows constant mean value $\langle \hat q \rangle =0$, and chacteristic fluctuations $\Delta^2q=1/2$. Coherent states $\rho=\vert\alpha\rangle\langle\alpha\vert$ exhibit oscillations in $\langle \hat q \rangle =\sqrt{2}\alpha\cos(\vartheta)$, while maintaining the same fluctuations as the vacuum $\Delta^2q=1/2$.}
    \label{fig:homodyne}
\end{figure}

Indeed, the values of the quadratures can be accessed in the experiment by means of a homodyne detector~\cite{Bachor}, shown in Fig.~\ref{fig:homodyne}: the light on our mode is made interfere with an intense local oscillator on a beamsplitter with equal reflectivity and transmittivity. Light at the two outputs reaches linear detectors and the difference of the two photocurrents can be demonstrated to be proportional to $|\alpha_{\rm LO}|(e^{i\vartheta}\hat a+e^{-i\vartheta}\hat a^\dag)$,
where $\alpha_{\rm LO}=|\alpha_{\rm LO}|e^{i\vartheta}$ is the complex amplitude of the local oscillator. This operation thus maps exactly the statistics of a generalised quadrature $\hat q_\vartheta = \cos\vartheta \hat x+ \sin \vartheta \hat p$ on that of the output current, with an amplification factor $|\alpha_{\rm LO}|$. There is an unknown conversion factor between the measured current and the actual values of the quadratures, but this can be directly inferred by imposing that the variance of the current with a vacuum input is 1/2. By tuning the phase of the local oscillator $\vartheta$, we can set whether the quadrature $\hat x$ is measured ($\vartheta=0$), or $\hat p$ ($\vartheta=\pi/2$), or any generalised quadrature in between. 

Differently from photon counting, homodyne detection is mode sensitive: detection occurs only for the mode that perfectly matches the local oscillator on the beamsplitter. Any difference between these two modes, leading to an estimated interference fringe visibility $v$, would decrease the effective efficiency of the detection by a factor $v^2$~\cite{refId0}. A second crucial aspect in homodyne detection is achieving the balance between the two outputs of the beamsplitter, which must be as equal as possible. Indeed, subtraction of the two currents is necessary to cancel classical noise on the local oscillator, however unavoidable discrepancies will make it show in the final signal. The better the balance, the higher the value of $|\alpha_{\rm LO}|$ which can be employed, for the benefit of higher rejection of the electronic noise from the linear detectors.  

\subsection{Fisher information and the Cram\'er-Rao bound}

Turning back to our original problem, we have developed an intuition that a measure of information is an appropriate figure to associate to different strategies, but we are also left with the problem of connecting it to the uncertainty on our parameters. This problem is not necessarily quantum, since classical measurement strategies work by the same conceptual scheme, only subject to different limitations. We can then start looking into solutions of this issue at the classical level, and then extend them to the quantum case. We will focus on {\it local estimation} of the parameters. By this, we mean that we know approximately their values, thanks, for instance, to preliminary coarse measurements we now wish to refine. Each measurement run will give a measured value $x_i$ for a quantity we know to be somehow related to a parameter of interest $\phi$; this could be, in principle, a vector of parameters, but we should better discuss the single-parameter case first. 

The probability distribution of the measured values is then $p(x|\phi)$, which we can interpret as the conditional probability of observing the value $x$, given that the parameter assumes the value $\phi$. In most scenarios, we are able to collect the outcomes of $M$ repeated experiments: $\{x_1,x_2,...,x_M\}$, all drawn from $p(x|\phi)$. Since we are assuming complete knowledge on how the parameter $\phi$ enters in the expression of $p(x|\phi)$, this provide us with means to give an estimation $\tilde \phi$ of the value of the parameter. We have at our disposal a function mapping the measured values $\{x_i\}$ to a value  $\tilde \phi(\{x_i\})$: such a function is called an {\it estimator}. The quality of different estimators of $\phi$ will depend on the measured quantity, and the final result will also be influenced on the number of measurements. Two quantities need to be inspected: how close the value of our estimator is to the actual parameter, and how wide its distribution is on average. The first quantity is the bias $b={\bf E}\left(\tilde \phi(\{x_i\})-\phi\right)$ -- here $\bf{E}$ denotes the expectation value on all possible outcomes $\{x_i\}$ for a given $\phi$. We are interested in {\it unbiased estimators}, {\it i.e.} $b=0$ for every $\phi$, giving the value of the parameter with arbitrary accuracy. For such unbiased estimators the variance 
$\sigma^2={\bf E}\left((\tilde \phi(\{x_i\})-\phi)^2\right)$ gives indications on the precision.

In order to compare different estimation strategies in quantitative terms,
we first define the {\it score} as~\cite{doi:10.1098/rsta.1922.0009}
\begin{equation}
    \label{score}
    V(x,\phi)=\frac{\partial \log p(x|\phi)}{\partial \phi}.
\end{equation}
This quantity indicates the relative variation of the probability of the measured value $x$ when the parameter $\phi$ undergoes slight changes. Under certain regularity conditions, it can be demonstrated that its expectation value, taken over all possible outcomes $x$, vanishes. We thus turn our attention to its variance, going under the name of {\it Fisher information}~\cite{doi:10.1098/rsta.1922.0009}:
\begin{equation}
\begin{aligned}
    \label{fisherclassica}
    F(\phi)= & \mathbf{E}[V(x,\phi)^2]\\
    = & \int~dx~p(x|\phi) \left(\frac{\partial \log p(x|\phi)}{\partial \phi} \right)^2,\\
    = & \int~dx~\frac{1}{p(x|\phi)} \left(\frac{\partial p(x|\phi)}{\partial \phi} \right)^2.
\end{aligned}
\end{equation}
This quantity is then linked to how a change in the value of the parameter affects, on average, relative variations in the probability distribution of the measured quantity $x$. For unbiased estimators, an inequality can be found 
\begin{equation}
\label{ccrb}
    \sigma^2\geq \frac{1}{M F(\phi)},
\end{equation}
which is the celebrated Cram\'er-Rao bound (CRB) \cite{Cramer, Rao}. Details on its derivation are discussed in Appendix B. This inequality sets the minimal variance attainable by repeating the experiment $M$ times, and by processing the data by means of a proper estimator. This presumes that the measurements are affected only by the fluctuations due to the outcome statistics $p(x|\phi)$ and no more.  If we are confident that out sample is sufficiently large, a variance exceeding the minimum in \eqref{ccrb} reveals the presence of some kind of noise, not explicitly considered in the outcome distribution, either arising from technical limitations, or large variations of the parameter $\phi$ itself.  

\subsection{Estimators}
There exist different options to data processing that deliver an unbiased estimator, able to saturate, in principle, the CRB. The simpler choice is to build a maximum likelihood estimator, by considering that, for independent runs, the probability of observing a specific collection of outcomes $\{x_1,..,x_M\}$ is
\begin{equation}
    \mathcal{L}(\{x_i\}|\phi)= \prod_{i=1}^M p(x_i|\phi),
\end{equation}
which we can take as a likelihood function. The estimated value of $\phi$ can thus be taken as the one maximising the likelihood:
\begin{equation}
\label{maxlik}
    \tilde \phi = \text{arg} \max_\phi\, \mathcal{L}(\{x_i\}|\phi).
\end{equation}
An uncertainty on $\tilde \phi$ is assessed by either repeating the set of $M$ runs multiple times, or, if one is reasonably confident of the outcome statistics, by applying a bootstrap method to the data in order to simulate further experiments by a Monte Carlo routine. This procedure is preferable to trying and applying error propagation to \eqref{maxlik}, since it automatically takes into account correlations within the data. 

An alternative method is grounded in Bayesian analysis, and considers the parameter $\phi$ itself as a statistical variable. The likelihood function should then satisfy Bayes' rule, leading to the expression:
\begin{equation}
\label{bayes}
     P(\phi|\{x_i\})= \mathcal{L}(\{x_i\}|\phi) P(\phi)/ P(\{x_i\}),
\end{equation}
where $P(\phi)$ is the a priori probability for $\phi$ - which we have assumed to be narrowly distributed - and $P(\phi|\{x_i\})$ is the updated conditional probability for $\phi$, given the observed $\{x_i\}$. Finally, $P(\{x_i\})$ is the probability of the experimental outcome, which we can calculate by normalising the conditional probability. The estimate of the parameter and of its uncertainty are thus assessed by calculating the first and second moments of \eqref{bayes}.

These two estimators are known to be optimal, in that they saturate the CRB~\cite{Rao}; however, for this condition to be met, the asymptotic regime of a large collection of runs is needed. In the practice, this limit can typically be reached with $M\simeq1000$~\cite{PhysRevLett.76.4295, PhysRevA.85.043817}. Thus, even if the CRB holds for arbitrary $M$, with standard estimators we expect this to provide useful guidance only in this regime. When only small samples are available, saturating the CRB is still possible, but requires more sophisticated data processing~\cite{PhysRevLett.117.010503,PhysRevApplied.10.044033,Rubio_2018}. A modified CRB can be derived also in the case of biased estimators~\cite{VanTrees}: $\sigma^2 \geq (1+db/d\phi)^2/F(\phi)$. If $db/d\phi<0$, the variance can reach a value below the unbiased CRB. Conversely, if the analysis reveals an uncertainty below the minimum, this can be a symptom of a biased estimator or the consequence of too small a collected sample.
When an experimental estimate of the variance $\sigma^2$ is obtained, it can be compared with the one at the CRB $\sigma^2_0=1/\left(M F(\phi)\right)$. For this purpose, the ratio $\mathcal{F}=\sigma^2/\sigma_0^2=M F(\phi) \sigma^2$ is considered, which is expected to be distributed according to the $\chi^2$ distribution.

\subsection{The Mach-Zehnder interferometer}
We could now use an example to help clarifying. The single-photon Mach-Zehnder interferometer (MZI) is an instructive choice, summarising all of the different aspects we have discussed. Incidentally, it is as relevant to optics as it is for atoms, since Ramsey interferometry can be directly translated into an equivalent MZI~\cite{PhysRev.78.695}.

\begin{figure}
    \centering
    \includegraphics[width=\columnwidth]{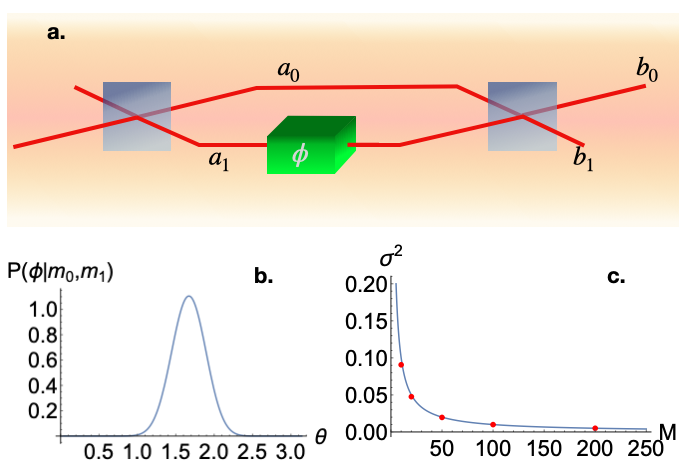}
    \caption{Phase estimation in a Mach-Zehnder interferometer. a) The basic scheme of the MZI - two mirrors and a phase shifter - with the labels for the input $a_0,a_1$ and output $b_0,b_1$ modes. b. A posteriori Bayesian probability distribution for the phase $\phi$, based on $M=20$ repetitions. c. Variance on the estimated phase as a function of $M$: the red points are numerical simulations, while the blue continuous curve shows the CRB $\sigma^2=1/M$.}
    \label{fig:MachZehender}
\end{figure}

The scheme, shown in Fig.~\ref{fig:MachZehender}a, has a single photon arriving onto a beam-splitter with no absorption, and whose tramsittivity equals its reflectivity ($|r|^2=|t|^2=1/2$); a relative phase shift $\phi$ occurs between the two output modes which are eventually recombined on a second beam-splitter. The phase $\phi$ is the parameter we want to estimate. Calling $\hat a_0^\dag$ and $\hat a_1^\dag$ the creation operators for the two output modes, the action of this phase shift is dictated by an operator $\hat U = \exp(i\frac{\phi}{2}(\hat a_1^\dag \hat a_1-\hat a_0^\dag \hat a_0))$. We can describe the quantum state of the light in two equivalent manners: the most intuitive one has the photon in a quantum superposition of being on either mode, while the more rigorous one considers the occupation number of the two modes: $\ket{\psi_\phi}=2^{-1/2}(\hat a_0^\dag+e^{i\phi}\hat a_1^\dag)\ket{0}$, where $\ket{0}$ is the common vacuum state. This latter view is the most convenient when it comes to generalising to more particles. The phase shift $\phi$ is retrieved by superposing the two modes $a_0$ and $a_1$ on a second beamsplitter, and then performing photon counting at two two outputs $b_0$ and $b_1$. The probabilities of finding a photon on either mode are $p_{0}(\phi)=\cos^2(\phi/2)$, and $p_{1}(\phi)=\sin^2(\phi/2)$. 

We thus recognise in this simple arrangement the three steps we have discussed above: the first beamsplitter prepares the single-photon (or, better, the two-mode) state as the probe; the sample determines the interaction, characterised by the parameter $\phi$, and, finally, the second beamsplitter and the photon counters represent the measurement stage. For this particular case, the limit on the precision on the phase $\phi$ is set by the Fisher information, found summing over the two possible outcomes $x=0$ and $x=1$:
\begin{equation}
\begin{aligned}
\label{fisingolo}
    F(\phi)=& \frac{1}{p_0}\left(\frac{\partial p_0}{\partial{\phi}}\right)^2+\frac{1}{p_1}\left(\frac{\partial p_1}{\partial{\phi}}\right)^2\\
    =& 1.
    \end{aligned}
\end{equation}
This implies that, by a series of $M$ repetitions, any phase can be estimated, in principle, with uncertainty $\sigma=1/\sqrt{M}$. For the purpose of estimating $\phi$, a Bayesian estimator can be employed: knowing that a number $m_0$ of events lead to a photon recorded on mode $b_0$, and $m_1=M-m_0$ on mode $b_1$, we can define the a posteriori probability distribution for $\phi$ as $P(\phi|m0,m1)=p_0(\phi)^{m_0}p_1(\phi)^{m_1} P(\phi)$, up to normalisation. This distribution is  shown in Fig.~\ref{fig:MachZehender}b for $M=20$. Its average value will deliver the estimate $\tilde \phi$, while its second moment is a measure of the uncertainty $\sigma^2$. Fig.~\ref{fig:MachZehender}c. illustrates the variance obtained in simulated experiments, for different values of $M$. Close inspection reveals that with a small number of repetitions, the variance of our estimator falls below the CRB. From $M\simeq 200$, proper working conditions are met, in line with previous considerations on the strictness of the CRB in the regime of a  large number of copies $M$. It should be noted that $P(\phi|m_0,m_1)$ with a flat prior may become a multimodal distribution, {\it i.e.} it shows different peaks. In this case, one can leverage on the fact that the CRB only holds for local estimation, and use the a priori information $P(\phi)$ to rule out irrelevant cases; in our example, we have employed a flat prior in the $[0,\pi]$ range. We are also assuming we can restrict the possible values of our phase well enough that we can ignore the fact they are actually circular variables, see {\it e.g.}~\cite{PhysRevLett.97.243601}.

Nothing prevents from using classical light in the MZI, and consider what the ultimate precision limit is. The analysis now considers a coherent state $\ket{\alpha}$, as a way of describing a classical field through quantum formalism. For the sake of comparing strategies with equal resources, we fix the average photon number in the coherent state as 
$|\alpha|^2=M$. The action of the MZI results in two coherent states $\ket{\alpha \cos(\phi/2)}$ and $\ket{\alpha \sin(\phi/2)}$ appearing on the two output modes: the intensity of the incoming beam is thus eventually parted among the two output modes as $I_0 = |\alpha|^2 \cos^2(\phi/2)$, and $I_1 = |\alpha|^2 \sin^2(\phi/2)$~\cite{Bachor}. The uncertainty on $\phi$ estimated from an intensity measurement on one of the arms is $\sigma =\left\vert d I_0/d\phi\right \vert^{-1}\Delta I_0$, where $\Delta I_0$ gives the size of the intensity fluctuations. For a coherent state, the variance of its photon number equals the average, therefore $\Delta I_0=\alpha^2\cos^2(\phi/2)$. This leads to the same expression for the uncertainty  $\sigma = 1/|\alpha|=1/\sqrt{M}$ as for single photons: for a given total energy, the limit for classical light is the same as the one for the use of independent photons. This should not come at a surprise: measurements of the photon number in a coherent state are described as a Poisson distribution with mean $|\alpha|^2$. This statistics is typical of independent events, thus it can not matter whether the photons are sent one by one in the MZI, or rather sent together, but acting independently. 
This is often called shot-noise limit (SNL) or standard quantum limit (SQL), albeit it is the one relevant for classical light~\cite{MandelWolf, Bachor,PhysRevD.23.1693}.

\section{Introducing quantum metrology}
\subsection{The quantum Cram\'er-Rao bound}
We now move back to the generic quantum case: after initialisation to a state $\rho_0$, the probe interacts with the sample, so that its state becomes $\rho_\phi$, and this is finally measured to extract the value of $\phi$ from the outcome distributions. We are confident about our {\it statistical model}: we know how $\rho_\phi$ is connected to the parameter $\phi$ for all instances. On the other hand, we can not be sure that our choice of initial probe state and final measurement is actually the most informative at our availability: after all, the probabilities $p(x|\phi)$ are obtained for a specific observable, but there are infinitely many other possibilities. 

The standard way of describing a measurement has it associated to an observable $\hat X$, with eigenvectors $\ket{x}$: the probability $p(x)$ of measuring the outcome $x$ is given by Born's rule $p(x)=\bra{x}\rho_\phi\ket{x}=\text{Tr}\left[\rho_\phi\ket{x}\bra{x}\right]$. Such measurements are called projective, as their action is captured by the projecting operator $\ket{x}\bra{x}$. This can be extended to more general cases that include imperfect measurements or non-discriminating observations: the outcomes are not associated to a canonical observable, but each value $x$ corresponds to an operator $\Pi_x$, such that Born's rule can still be written as $p(x)=\text{Tr}\left[\Pi_x\rho_\phi\Pi_x^\dag\right]$, and the outcomes $x$ cover all instances $\sum_x\Pi_x^\dag\Pi_x=\mathbb{I}$. These are useful tools in describing click detectors, and their inability of discriminating Fock states, as shown in Fig.~\ref{fig:detectors}.

To our purposes, the classical Fisher information \eqref{fisherclassica} provides a tool to compare different choices, but no constructive guideline. Intuitively, this is because the Fisher information looks at how the specific outcome distributions vary with the parameter $\phi$, rather than at the changes of the state itself. In order to establish such a notion on solid grounds, we first need to learn how to describe the derivative of a state in operatorial terms by introducing the symmetric logarithmic derivative (SLD) operator $L_\phi$ as~\cite{1054108}:
\begin{equation}
\label{SLD}
    \frac{\partial \rho_\phi}{\partial \phi} = \frac{1}{2}\left( L_\phi \rho_\phi+\rho_\phi L_\phi\right). 
\end{equation}
Indeed, it can be demonstrated that the quantity~\cite{1054108,1976iv,Holevo:1414149}
\begin{equation}
    H(\phi)=\text{Tr}\left[ L_\phi^2 \rho_\phi \right] 
\end{equation}
sets a higher bound on the Fisher information for all possible choices of the measurement~\cite{PhysRevLett.72.3439, Barndorff_Nielsen_2000}; further, a related result is that there always exist a measurement that saturates the inequality, $F(\phi)=H(\phi)$, and it corresponds to a projective measurement in the eigenbasis of $L_\phi$~\cite{1054108}; these aspects are briefly discussed in Appendix C. The optimal measurement may not be unique, nor necessarily easy to implement, but it is guaranteed to exist, and provides guidance to assess how well the chosen strategy is working. This is why $H(\phi)$ is called the {\it quantum Fisher information} associated to the state $\rho_\phi$. The achievable precision is thus limited from below as:
\begin{equation}
\label{qcrb}
    \sigma^2\geq \frac{1}{M H(\phi)},
\end{equation}
an inequality that goes under the name of quantum Cram\'er-Rao bound (QCRB).

The way an experiment is usually designed considers a state $\rho_0$ in a given set, chosen according to certain criteria; for instance, these may be imposed by experimental limitations, or from preliminary considerations on the form of the state. The quantum Fisher information is then calculated for the evolved states $\rho_\phi$, keeping it a function of the variables defining the set. These are then optimised to provide the probe giving the highest quantum Fisher information. 

We can use these considerations to inspect the single-photon MZI we discussed above. If we now allow for arbitrary transmission $t$ and reflection $r=\sqrt{1-t^2}$ coefficients, the state in the interferometer can be written as $\ket{\psi_\phi}=(t\hat a_0^\dag+re^{i\phi}\hat a_1^\dag)\ket{0}$, or, making the occupation numbers explicit, $\ket{\psi_\phi}=\left(t\ket{1,0} +re^{i\phi}\ket{0,1}\right)$. For pure states, the defining equation of the SLD \eqref{SLD} greatly simplifies and the solution $L_\phi =2\left( \ket{\partial_\phi \psi_\phi}\bra{\psi_\phi}+\ket{\psi_\phi}\bra{\partial_\phi \psi_\phi}\right)$ is readily found~\cite{FUJIWARA1995119}; this finally leads to the expression
\begin{equation}
\label{purestate}
    H(\phi)=4\left[\bra{\partial_\phi \psi_\phi}\partial_\phi \psi_\phi\rangle +(\bra{\partial_\phi \psi_\phi} \psi_\phi\rangle)^2\right].
\end{equation}
Since in our case $\ket{\partial_\phi \psi_\phi}=ire^{i\phi}\ket{0,1}$, the quantum Fisher information writes $H(\phi)=4r^2(1-r^2)$: the optimal choices corresponds to the symmetric case $t=r=1/\sqrt{2}$, indeed. Further, the strategy we have devised achieves the maximum value for the Fisher information, thus it is able, in principle, to saturate the QCRB. 

\subsection{The origin of quantum enhancement}
Quantum states can be shown to offer superior precision, based on very general considerations, as presented in ~\cite{PhysRevLett.96.010401}, at least for {\it unitary} parameters, those that are related to the action of a unitary $\hat U_\phi=e^{-i\phi \hat G}$, $\hat G$ being the generator of the transformation. Fairly frequently, this corresponds to the Hamiltonian of the system.

For a pure initial state $\rho_0=\vert \psi_0\rangle\langle \psi_0\vert$, the evolution writes $\vert \psi_\phi\rangle=\hat U_\phi \vert \psi_0 \rangle$; the derivative of the state is then $\vert \partial_\phi \psi_\phi\rangle = -i\hat G \vert \psi_\phi \rangle$. From the expression \eqref{purestate} for the QFI, we obtain
\begin{equation}
    H(\phi)=4\Delta^2 G,
\end{equation}
{\it i.e.} the QFI equals the variance of the generator $\hat G$ on the state $\vert \psi_\phi \rangle$. This gives the QCRB a form resembling Heisenberg's relation~\cite{BRAUNSTEIN1996135,PhysRevLett.98.160401}:
\begin{equation}
    \sigma^2 \,\Delta^2 G\geq \frac{1}{4M}.
\end{equation}

This expression helps us to draw a comparison between classical and quantum strategies for estimation. It is then clear that the aim for a classical and a quantum experimenter is to prepare a state maximizing the variance $\Delta^2G$. It may first seem contradictory to label as classical an experimenter who is given access to quantum states: however, if their capabilities are cunningly restricted, we can obtain bounds pertaining to classical resources; the example of the Mach-Zehender interferometer is quite illustrative. 

 We allow both the classical and the quantum laboratories to use $N$ particles for each experimental run. The classical experimenter can thus only prepare the state of individual particles in the equal superposition of eigenstates of the maximal $g_M$ and minimal $g_m$ eigenvalues of $\hat G$: $\vert \psi_0\rangle=(\vert g_M\rangle +\vert g_m \rangle)/\sqrt{2}$. Since in this state $\Delta^2G = (g_M-g_m)^2/4$ and the run is using $N$ such states, the total QFI is $H(\phi)=N(g_M-g_m)^2$, growing linearly with $N$. This describes exactly what happens in the MZI, thus we recognise this scaling as the optimal classical limit, {i.e} the SQL. The quantum experimenter has more freedom, and can prepare collective states of all $N$ particles; in particular, if the global generator $\hat G^{\otimes N}$ is considered, the experimenter can prepare a superposition of global eigenstates $\vert \psi_0 \rangle=(\vert g_M\rangle^{\otimes N} +\vert g_m \rangle^{\otimes N})/\sqrt{2}$, that exhibits a variance, hence a QFI,  $H(
\phi) = N^2(g_M-g_m)^2$. The scaling of the QFI is improved with respect to the classical case~\cite{PhysRevA.33.4033}, and takes the name of Heisenberg limit (HL)~\cite{PhysRevA.55.2598}. In principle, the scaling at the SQL or the HL can be achieved with separable measurements, if we can retain control on individual particles~\cite{PhysRevLett.96.010401}, while in the other instances collective measurements are required.

Some considerations on the use of the resources are in order: the whole experiment comprises multiple runs, $M$, thus $N\cdot M$ particles are used overall. Thus, the view that SQL has the QFI scaling linearly, and the HL scaling quadratically with the number of resources glosses over this observation, but it captures correctly what happens in each single run. In the same vein, a single particle may still provide a similar advantage, if it is used to investigate the sample $N$ times, by implementing the transformation $U_\phi^N$: here, the scaling of the QFI is again quadratic in $N$, and has also been described as HL~\cite{HLNature,Higgins_2009,HLvero}, with each passage counting as one resource in a run. This justifies the comment above on the disparate nature of resources in quantum metrology. We notice that, stretching beyond this simple unitary case, the use of Hamiltonians with $k$-body interactions has been shown to provide $N^{-k}$ scaling for the variance~\cite{PhysRevA.77.012317,Napolitano}.

\section{Applications}
We have thus learned how precious Fisher information is when we need to assess strategies for parameter estimation. We should not forget, on the other hand, that this serves one purpose: guiding the design of resource states and their measurement. We will now discuss  relevant examples centered on the estimation of optical phases. 

\subsection{Optical interferometry with squeezed states}

A phase in an optical interferometer is the most common example studied in quantum estimation, due to its conceptual value, as well as to its relevance to many experimental situations. Certainly, it is indebted for much of its popularity to the connection to the measurement of gravitational waves by long-arm interferometers~\cite{ABBOTT2004154,Abramovici325}. One of the most influential results in this field is the work in~\cite{PhysRevD.23.1693} that proposed the use of squeezed state as a way to reduce noise in these interferometers.

We start inspecting our classical MZI under a different perspective, so that we can now include quantum fluctuations more explicitly. We write the output operators $\hat b_0$ and $\hat b_1$ in terms of the input operators, referring to the modes arriving onto the first beamsplitter, which we call $\hat i_0$ and $\hat i_1$. Up to overall phases, these relations write $\hat b_0 = \cos(\phi/2)\hat i_0 +\sin(\phi/2)\hat i_1 $, and  $\hat b_1 = \cos(\phi/2)\hat i_1 -\sin(\phi/2)\hat i_0$~\cite{walls1995quantum,Loudon:105699}. The intensity on the arm $b_0$ is then given by: $\hat b_0^\dag \hat b_0=\cos^2(\phi/2) \hat i_0^\dag \hat i_0+\sin^2(\phi/2) \hat i_1^\dag \hat i_1 +\sin(\phi)/2(\hat i_0^\dag \hat i_1+\hat i_0 \hat i_1^\dag)$. The standard arrangement has an intense beam with classical amplitude $\alpha$, taken to be real, on the input $i_0$, and vacuum on $i_1$: we can thus replace the operators $\hat i_0$ and $\hat i_0^\dag$ with the corresponding classical numbers. Consequently, the intensity operator becomes: $\hat b_0^\dag \hat b_0=\cos^2(\phi/2) \alpha^2 +\left(\sin(\phi)\alpha/\sqrt{2}\right)\hat x$, where $\hat x$ is the $x$ quadrature \emph{of the vacuum mode} $i_1$. This treatment is consistent with the approach we have taken in the previous section, and leads to the correct expression $\sigma^2=1/\alpha^2$ for the uncertainty on the phase. More relevantly to our purposes, this expression offers an intriguing picture: the shot noise observed with coherent states is a consequence of the fluctuations of the vacuum modes entering the apparatus through the unused port, as originally recognised in~\cite{PhysRevLett.45.75}.

The only way of preventing vacuum from entering is replacing it with another state: improving on the shot noise then demands controlling the fluctuations. In fact, the variances on the quadrature operators ought to satisfy Heisenberg's relation $\Delta^2 x \Delta^2 p\geq 1/4$, properly rescaled to our choice of the units, yet there is no individual bound on either; we can reduce one at will, provided the other is increased consistently. Therefore, a suppression of the fluctuations on the $\hat x$ quadrature as $\Delta^2 x= e^{-2s}/2$ must be compensated by increased fluctuations on the $\hat p$ quadrature by at least $\Delta^2 p= e^{2s}/2$. State displaying such a property are called squeezed states~\cite{1965Tgso, PhysRevA.13.2226, PhysRevD.23.1693,squeezedreview}; in particular, when the quadratures are centred on zero, the state takes the name of squeezed vacuum, although the average photon number in the squeezed vacuum is not zero, but $\bar n = \sinh^2(s)$~\cite{Loudon:105699}.  Thanks to squeezing, the uncertainty in our phase measurement can now be improved as $\sigma^2 = e^{-2s}/\alpha^2$, in the regime $\bar n\ll\vert\alpha\vert^2$  ~\cite{PhysRevD.23.1693}. 
This scheme already offers a concrete advantage, nevertheless, it is not optimal: an explicit calculation gives $F(\phi)=\alpha^2e^{-2s}+\bar n$~\cite{PhysRevLett.100.073601}. The reason why this result is not recovered by means of error propagation can be traced to the poor performance of the average intensity as the estimator~\cite{PhysRevLett.100.073601}. If the energy is equally parted between the coherent state and the squeezed state, {\it i.e.} $|\alpha|^2=\bar n$, this implies HL with the total number of photons.

\begin{figure}
    \centering
    \includegraphics[width=\columnwidth]{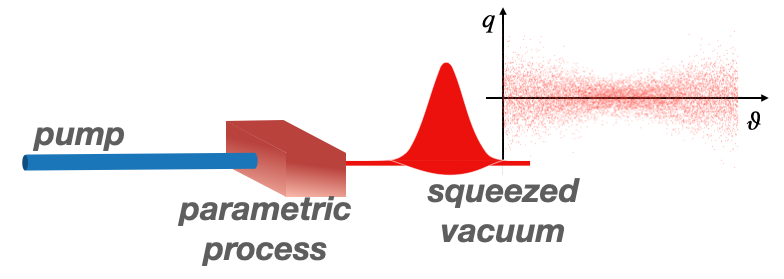}
    \caption{Scheme for the production of squeezed vacuum states. A pump beam provide the energy to drive a nonlinear optical parametric process by which squeezed vacuum is produced. The Hamiltonian pertaining to this coupling is quadratic in $\hat a$ and $\hat a^\dag$, ensuring the even-photon number superposition in~\eqref{eq:squeezed}. The variance of the quadrature $\hat q$ thus depends on the phase $\vartheta$.}
    \label{fig:squeezing}
\end{figure}

Producing squeezed vacuum in a laboratory requires processes in which photons are produced in pairs, since its expression in Fock states writes~\cite{Loudon:105699}
\begin{equation}
\label{eq:squeezed}
    \vert \zeta \rangle = \text{sech}^{1/2}s \sum_{n=0}^{\infty}\frac{[2n!]^{1/2}}{n!}\left[-\frac{1}{2}e^{i\vartheta}\tanh s\right]^n\vert2n\rangle,
\end{equation}
where $\vartheta$ is associated to the squeezed quadrature. Nonlinear optics offers a solution by means of parametric processes, in which two photons originate from the conversion of one photon in an optical crystal (parametric downconversion), or two photons in atomic vapours or fibres (four-wave mixing), as depicted in Fig.~\ref{fig:squeezing}. The first reported production of squeezed light is the one in~\cite{PhysRevLett.55.2409}: the source was based on four-wave mixing in a sodium gas jet in an optical a cavity, and a noise reduction of 0.3dB was achieved. This was followed shortly by other experiments based on optical nonlinearities in a fibre~\cite{PhysRevLett.57.691} and in a crystal~\cite{PhysRevLett.57.2520}.  Early examples of the enhancement in precision based on squeezed light are found in~\cite{PhysRevLett.59.278, PhysRevLett.59.2153}.

The sources are often inserted in a cavity in order to enhance the nonlinear optical interaction with the pump beam: squeezing is thus revealed in the noise reduction in the sidebands of a continuous homodyne signal~\cite{Bachor}. These are normally operated at a few MHz~\cite{PhysRevLett.100.033602}, but, thanks to specific technical solutions, operation in the GHz regime can also be achieved~\cite{Senior:07,Ast:13}. Extending the working range down to the smaller frequencies is much more challenging, in comparison, due to the presence of classical noise in that region: it took decades of patient craftsmanship~\cite{GWreview} in order to make it possible to put squeezing at the service of gravitational wave detection~\cite{LIGO}. Alternatively, pulses can be adopted for pumping, making it possible to produce squeezing in the time domain~\cite{PhysRevLett.59.2566,Wenger:04,Eto:07}. Finally, these two approaches can be merged in order to produce squeezing in frequency combs~\cite{PhysRevLett.108.083601}. 

If there is a Moriarty to every Holmes, loss does indeed play that role for quantum enhancement: some of the photons do not contribute to the  final signal, nonetheless these resources have been produced and prepared, but wasted for the sake of estimation. This is a severe but not ruinous restriction when using squeezed states: if  transmission occurs with loss $1-\eta$, the variance of the squeezed quadrature will be a weighted sum of the initial one and that of the vacuum: $\Delta^2 x = \eta e^{-2s}/2 +(1-\eta)/2.$  Noise suppression is reduced - the vacuum has found its way back into the interferometer - but not lost. This mechanism explains why squeezing is the optimal choice for gravitational wave measurements~\cite{PhysRevA.88.041802}; currently it has been applied with success to more involved problems, including monitoring of biological specimens~\cite{biosqueezing}, tracking of time-changing phases~\cite{Yonezawa1514}, and problems in magnetometry~\cite{PhysRevLett.105.053601,Li:18}. On the other hand, the detection scheme is sensibly affected by phase fluctuations. This amounts to averaging the noise over rotated quadratures, which have components along $x$, the squeezed direction, as well as $p$, which has excess noise with respect to the SNL: this noise will enter the detection through this averaging, hence spoiling the enhancement. Such a mechanism sets practical limits to the amount of squeezing that may be efficiently employed. This is made even worse by the fact that the pure squeezed vacuum \eqref{eq:squeezed} is a distant approximation of the state actually produced: parasitic nonlinear processes couple the squeezed mode to others, resulting in excess noise in the $\hat p$ quadrature, with respect to what is expected from the level of squeezing~\cite{Ourjoumtsev83}.

For problems involving two modes, the use of entangled two-mode squeezed vacuum can be relevant; these are produced by interference of two squeezed vacuum modes on a symmetric beam splitter, with the same level of squeezing, but along orthogonal directions in phase space.  They show correlations in the value of the quadratures of the two modes, {\it e.g.} $p_1$ and $p_2$, in that the variance of their difference $\Delta^2\left((p_1-p_2)/\sqrt{2}\right)$ remains below the vacuum noise level~\cite{Loudon:105699,PhysRevLett.59.2555}. This implies that the conjugate quadrature $(x_1-x_2)/\sqrt{2}$ must show increased fluctuations. The other linear combination $\left((x_1+x_2)/\sqrt{2}\right)$, instead, is squeezed by the same amount. These states can also be produced by means of a nonlinear optical interaction, either parametric downconversion or four-wave mixing, realised by coupling the pump mode to two modes at lower frequencies, made distinguishable in direction or mean wavelength. A suppression of the variance by a factor $e^{-2s}$ corresponds to the state:
\begin{equation}
\label{eq:tmss}
    \vert \zeta_2 \rangle = \frac{1}{\cosh s}\sum_{n=0}^{\infty} \left[-e^{i\vartheta}\tanh s\right]^n \ket{n}_1\ket{n}_2,
\end{equation}
where, as above, $\vartheta$ identifies a pair of squeezed quadratures. The photon numbers of the two modes are perfectly correlated, an effect at the basis of their application to quantum imaging~\cite{PhysRevLett.59.2555,PhysRevLett.88.203601,Brida:2010jk,Boyer544}, and quantum plasmonic sensing~\cite{Dowran:18}. Their usefulness in correlated interferometry, in which two correlated phases pertain to two distinct MZIs, has also been demonstrated~\cite{Pradyumna:2020pd}. As with their single-mode sisters, ideal two-mode squeezed states in~\eqref{eq:tmss} are an idealisation of an experimental case degraded by loss and parasitic processes; nevertheless, they provide good guidance in experimental design.

\subsection{Fixed photon-number states}
\begin{figure}
    \centering
    \includegraphics[width=\columnwidth]{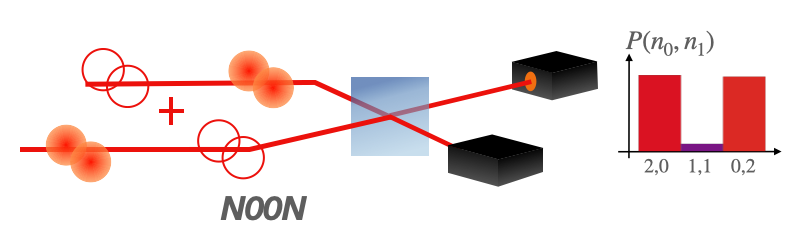}
    \caption{Illustration of $N00N$ states, here with $N=2$. In the optimal measurement scheme, the two modes are superposed on a beam splitter, before resolving the photon number on the two outputs: the phase $\phi$ can be reconstructed based on the observed frequencies of the different occupation numbers.}
    \label{fig:noon}
\end{figure}

When using electromagnetic fields, we can find it convenient to design the state fixing the number $N$ of photons that can be used in a run. For phase estimation these need to be subdivided between a probing arm, on which our target object sits, and a reference arm; the state is thus written in the most general form as:
\begin{equation}
    \label{eq:fixedN}
    \begin{aligned}
    \vert \psi \rangle = & \sum_{k=0}^N \alpha_k \left(\hat a_p^\dag\right)^k  \left(\hat a_r^\dag\right)^{N-k}\vert 0 \rangle\\
    = & \sum_{k=0}^N \beta_k \vert k \rangle\vert N-k \rangle,
\end{aligned}
\end{equation}
where $\hat a_p^\dag$ ($\hat a_r^\dag$) is the creation operator for the probing (reference) mode. The coefficients $\alpha_k$ or, equivalently, $\beta_k$ are chosen in order to maximise the Fisher information for a given evolution. In the simplest case, the phase is imparted by the operator $\hat U(\phi)=e^{i\phi(\hat n_p- \hat n_r)/2}$, as in the MZI. The optimal state is thus the one with the highest variance $\Delta^2(n_p-n_r)$ corresponding to $\beta_0=\beta_N=1/\sqrt{2}$, and the remaining coefficients being zero, as first recognised in~\cite{PhysRevLett.56.1515}, see Fig.~\ref{fig:noon}. These states are commonly called $N00N$ states, with an obvious pun on their expression in the Fock basis~\cite{RosettaStone}, and provide a QFI $H(\phi)=N^2$, reaching Heisenberg scaling in line with the considerations in the previous chapter. It should be noticed that we have not control on individual photons in these states: in a particle picture, a collective measurement on all $N$ photons is required for optimal extraction of information.

$N00N$ states have been the workhorse for quantum metrological tasks in many proof-of-principle experiments~\cite{PhysRevLett.65.1348,PhysRevLett.87.013602,PhysRevLett.89.213601,PhysRevLett.89.213601,doi:10.1063/1.4724105,photonicgear,delicate,Ono}. A common device for their production, which only works for $N=2$, is to make two indistinguishable photons arrive on the same time at a beam splitter~\cite{PhysRevLett.59.2044}: the output state is in the $N00N$ form $\vert20\rangle+\vert02\rangle$ over the two output modes. Here, indistinguishability is key to ensure that the two-photon interference process suppresses the $\vert 11\rangle$ component~\cite{branczyk2017hongoumandel,Bouchard_2021}, which would not contribute to our measurement task; when used in the Mach-Zehnder configuration, this term would lead to fringes with reduced visibility. Extending this to the interference of two Fock states does not lead to generic $N00N$, nevertheless the output state, often going under the name of Holland-Burnett state, is again capable of providing a $1/N^2$ scaling of the variance~\cite{PhysRevLett.71.1355}. 

Although there exist nowadays multiple solutions for producing single photons~\cite{doi:10.1063/1.3610677,ReviewSP}, a recipe for using them to build $N00N$ states deterministically is not known, let alone arbitrary states; there are, nevertheless, different proposals for their probabilistic, but heralded, production~\cite{PhysRevA.65.052104,PhysRevA.68.052315,PhysRevLett.99.163604,PhysRevA.76.031806}. Ad hoc solutions have been devised in order to produce states with $N>2$~\cite{mitchellnoon,PhysRevLett.107.163602,Afek879}, or states with similar metrological power~\cite{walthernoon,Nagata726,PhysRevLett.117.210502,PhysRevLett.120.260502,WalmsleyN}.

It was emphasised, from the very first reports~\cite{mitchellnoon}, that the presence of fringes oscillating with $N\phi$, what is called super-resolution, does not guarantee {\it per se} super-sensitivity, {\it i.e.} improved uncertainty. In fact, the super-resolution mechanism can be, and has been~\cite{PhysRevLett.98.223601,PhysRevA.81.063836}, replicated with classical light: it just takes to measure coincidences in a clever interferometer arrangement. When compared with squeezed states, loss has a more harmful impact on quantum phase estimation performed with $N00N$ states, and, in general, with fixed-photon states: with these probes, the advantage in using quantum resources may be completely spoilt. This can be understood as follows: a coherent state with $N$ photons on average will be transformed by loss to a different coherent state with lower energy~\cite{Loudon:105699}. The associated minimal uncertainty is then increased to $1/\sqrt{\eta N}$. In a $N00N$ state, instead, the loss of a single photon destroys quantum coherence - one would be able to tell from which mode the photon came - thus, only those events leading to all $N$ photons being detected provide information on the phase; the probability of this event scales as $\eta^{N}$, reducing the available Fisher information accordingly~\cite{PhysRevLett.102.040403}. Using a $N00N$ state is not necessarily the wisest choice, and optimal coefficients $\beta_k$ in \eqref{eq:fixedN} can be calculated for the actual lossy evolution~\cite{PhysRevLett.102.040403,PhysRevA.83.021804,PhysRevA.83.063836}. The expressions are rather cumbersome, and it is hard to derive general considerations other than the higher the loss, the more complex the structure of the state. Since the values of the coefficients depend on the value of $\eta$, it needs to be known in advance in order to design the state; an experimental realisation has been reported in~\cite{konrad}, with further work emphasising the possibility of remedying, in part, at the measurement stage~\cite{PhysRevLett.108.233602}. However, it may not be possible or convenient to design a sensor that can tailor states to the channel: fortunately, the class of Holland-Burnett states exhibit a good degree of resistance to loss, and can be considered as practical resources for lossy quantum phase estimation~\cite{PhysRevA.83.063836}. While a rigorous assessment of quantum advantage should consider a comparison of quantum and classical Fisher information, a simple rule writes~\cite{PhysRevLett.98.223601,Matthews16}
\begin{equation}
    \eta_{\rm tot} v^2 N>1,
\end{equation}
where $\eta_{\rm tot}$ is the total efficiency, including detection and generation when non-deterministic, and $v$ is the contrast of the fringes. Thanks to the enormous progress in high-efficiency detectors, demonstrations of an unconditional advantage in quantum phase estimation with two-photon $N00N$ states has been reported~\cite{phestforreal}. As far as scaling is concerned, however, the possibility of achieving the HL is compromised by decoherence~\cite{davidovich,elusive}. At best, we can realistically expect conditions sitting in between SNL and HL, unless we are able to access that part of the environment that is causing the noise: in that limit, a feedback mechanism can be implemented to recover part of the initial advantage~\cite{PhysRevLett.125.200505}. 

The working conditions for the interferometer should be set where sensitivity is the highest and more robust against imperfections; while, in ideal conditions, all values of the phase $\phi$ should show the same Fisher information, as exemplified by \eqref{fisingolo}, imperfect fringe visibility makes it convenient to fix $\phi$ to one specific value~\cite{PhysRevA.81.012305}. In gravitational wave detection, for instance, the interferometer operates close to the dark fringe condition~\cite{LIGO}, while in a MZI $\phi\simeq\pi/2$ is often preferred. Such an operation can be achieved by means of adaptive estimation~\cite{PhysRevLett.85.5098,PhysRevLett.89.133602}, which is typically based on Bayesian techniques; this has been demonstrated in~\cite{HLNature,PhysRevResearch.2.033078} with single photons, in~\cite{pryde} with $N00N$ states, and in \cite{phaseparis} with squeezing. In this case the CRB can not be applied to the whole estimation process, since it assumes local conditions upfront.

\section{Multiparameter estimation}

\subsection{The quantum Fisher information matrix}
This far, we have not looked into the fact that the probe state will emerge modified after the measurement, and how this translates in our context. Turning back to Heisenberg's example, we can observe that, first, the measurement is insensitive to the electron's momentum and, second, the state after the measurement carries no information on the original position and momentum. Therefore, if we aimed at collecting information on both quantities at once, oblivious of quantum mechanics, we would have failed spectacularly. 

Based on these considerations, we can now face the problem of generalising our treatment to the multiparameter case; we now aim at measuring a set of parameters $\vec\phi=\{ \phi_1, \phi_2,...,\phi_P\}$, searching to come as close as possible to the best possible precision. The values of the parameters are all inferred from the outcomes of $M$ repetitions of the measurement of the quantity $x$, exactly as in the single parameter case. This task is not as severely restricted as attempting the measurement of incompatible observables: we can design a strategy to end up with a value for each $\phi_i$. The limitations concern the quality of the estimation: the optimal measurement for one parameter may not be sensitive to some of the others or two or more such measurements can not be implemented jointly~\cite{PhysRevA.94.052108}. 

In order to describe uncertainty in the multiparameter case, the covariance matrix $\bf{\Sigma}$ is introduced:
\begin{equation}
    \Sigma_{h,k}={\bf E}\left((\tilde \phi_h-\phi_h)(\tilde \phi_k-\phi_k)\right).
\end{equation}
The diagonal element $\Sigma_{h,h}$ gives the variance on $\phi_h$, while the off diagonal elements satisfy $\Sigma_{h,k}=\Sigma_{k,h}$, and quantify how much our estimates of $\phi_h$ and $\phi_k$ are statistically correlated. 

The classical Fisher information itself can be expressed as a symmetric matrix~\cite{1976iv}:
\begin{equation}
\label{fim}
    F_{h,k}(\vec \phi)=\int dx \frac{1}{p(x|\vec \phi)}\frac{\partial p(x,\vec \phi)}{\partial \phi_h}\frac{\partial p(x,\vec \phi)}{\partial \phi_k}.
\end{equation}
The corresponding CRB then writes:
\begin{equation}
    {\bf \Sigma}\geq \frac{1}{M}{\bf F}^{-1},
\end{equation}
meaning that $\left({\bf \Sigma}-{\bf F}^{-1}/M\right)$ is a non-negative matrix or, equivalently, that for any unit vector $\vec u$ we have ${\vec u\cdot{\bf \Sigma} \vec u \geq \vec u\cdot {\bf F}^{-1} \vec u/M}$. In scalar form, we can use it to bound the individual variances as:
\begin{equation}
    \Sigma_{h,h}\geq \frac{1}{M}\left( F^{-1}\right)_{h,h}.
\end{equation}

We will now try and attach a meaning to both diagonal and off-diagonal terms in $\bf{F}$; for the sake of clarity, we consider the example of $P=2$ parameters, but our considerations extend to the general case. For these $2\times2$ matrices, the scalar bound for $\phi_1$ is given by the simple formula:
\begin{equation}
\Sigma_{1,1}\geq \frac{1}{ M}\frac{1}{F_{1,1}-F_{1,2}^2/F_{2,2}},
\end{equation}
and a similar expression holds for $\Sigma_{2,2}$. The scalar limit $\Sigma_{1,1}\geq 1/(M F_{1,1})$ is recovered when the diagonal term $F_{1,2}$ vanishes or when $F_{2,2}$ is infinite,{\it i.e.} when $\phi_2$ is perfectly known. This leads us to identify $F_{h,h}$ as the Fisher information on one parameter when all others are known, and $F_{h,k}$ as the quantifier of how much the uncertainty on the parameter $\phi_k$ affects that on $\phi_h$, and the other way around. The quantity $F^{(\rm eff)}_{1,1}=F_{1,1}-F_{1,2}^2/F_{2,2}$ represents an effective value for the available FI on the parameter $\phi_1$.

As for the single-parameter case, a comparison can be carried out between the experimental covariance matrix $\bf{\Sigma}$, and the prediction of the CRB ${\bf \Sigma}_0={\bf F}^{-1}/M$. Also in this case, a quantitative assessment can be carried out by means of a variable which is $\chi^2$-distributed~\cite{multivariate}. This has to take into account that, for the two matrices to match, both the magnitude of the uncertainties and their correlations must be compatible with the predictions.

\subsection{Using quantum resources}

The extension to the quantum case should be apparently straightforward: if one builds on the ideas leading to \eqref{fim}, it is natural to define a Quantum Fisher information matrix as~\cite{1976iv,Holevo:1414149}
\begin{equation}
\label{eq:matrixQFI}
    H_{h,k}(\vec \phi)=\frac{1}{2}\text{Tr}\left[\rho_{\vec \phi}\{L_h,L_k\}\right],
\end{equation}
where $L_h$ is the SLD associated to the parameter $\phi_h$, and the curly brackets denote the commutator. It can be demonstrated that a matrix QCRB holds:
\begin{equation}
\label{eq:MQCRB}
    {\bf \Sigma}\geq \frac{1}{M}{\bf F}^{-1}\geq \frac{1}{M}{\bf H}^{-1}.
\end{equation}
However, we are left without unambiguous guidance towards the optimal measurement, as the bound only tells us that each $L_h$ is connected to the best choice for $\phi_h$. More information is collected by looking at the so-called weak commutators
\begin{equation}
\label{eq:commSLD}
    D_{h,k}=\frac{i}{2}\text{Tr}\left[\rho_{\vec \phi}\left[L_h,L_k\right]\right].
\end{equation}
Whenever $D_{h,k}=0$, there exist a measurement able to achieve optimal precision for $\phi_h$ and $\phi_k$ at once;  unfortunately, the optimal choice may be a collective measurement performed on all $M$ copies~\cite{e21070703,PhysRevA.94.052108}.
We can still make use of the quantum Fisher information matrix to assess how well a given strategy is scoring, but, for this purpose, it is necessary to introduce scalar quantities to be compared. The usual choice is to consider a weighted sum of the individual variances $\Gamma=\text{Tr}[\bf{W\Sigma}^{-1}]$, where ${\bf W}$ is a diagonal matrix containing the weights; we thus obtain a lower limit
\begin{equation}
\label{eq:consld}
    \Gamma\geq\frac{1}{M}\text{Tr}[\bf{WH}^{-1}],
\end{equation}
and different strategies can be assessed on how close they come to it. A distinct evaluation criterion is based on assessing how much of the available information has actually been extracted by the selected measurement; the figure commonly employed for this purpose is
\begin{equation}
\label{eq:upsilon}
    \Upsilon(\vec \phi)=\text{Tr}\left[{\bf F}\,{\bf H}^{-1}\right].
\end{equation}
This quantity ranges from 0, in the trivial case of an uninformative measurement, to $P$, when all parameters are estimated jointly at their ultimate precision.

Should we be interested in a different parameter set $\vec\theta$, which is a function of the original $\vec \phi$, the corresponding Fisher information matrices, classical and quantum, are found as
\begin{equation}
\label{eq:repara}
\begin{aligned}
    {\bf F}_{\vec\theta}&={\bf B}\cdot{\bf F}_{\vec\phi}\cdot{\bf B}^{T},\\
    {\bf H}_{\vec\theta}&={\bf B}\cdot{\bf H}_{\vec\phi}\cdot{\bf B}^{T},
\end{aligned}
\end{equation}
where the elements of the reparametrisation matrix $\bf B$ are given by the derivatives $B_{i,j}=\partial \phi_j/\partial \theta_i$.

In a pure-state model, the evolution leads the initial state $\vert \psi_0\rangle$ to $\vert \psi_{\vec \phi}\rangle=e^{-i\vec G\cdot\vec \phi}\vert \psi_0\rangle$, where $\vec G$ is the vector of the generators associated to the different parameters. By a similar procedure as for the single-parameter case, we find
\begin{equation}
    \langle\psi_{\vec \phi}\vert L_hL_k\vert \psi_{\vec \phi}\rangle =4\left(\langle G_hG_k\rangle -\langle G_h\rangle\langle G_k\rangle\right), 
\end{equation}
therefore the Fisher information matrix~\eqref{eq:matrixQFI} is proportional to the symmetrised covariance matrix of the generators.

\subsection{Multiparameter Mach-Zehender interferometry}
As a much needed clarifying example, we consider a MZI in which we aim at estimating the phase $\phi$ and the transmittivity of the first BS $t$, jointly. As with our first example, we use a single photon as the input, while, for our measurement, we consider a second BS with variable transmittivity $t_m$, and one APD on each output mode. 

The state emerging from the evolution is in the form $\ket{\psi_{\phi,t}}=\left(t\ket{1,0} +re^{i\phi}\ket{0,1}\right)$ we discussed above, this time interpreted as a function of both parameters $\phi$ and $t$. The SLDs are found with the pure state model, and deliver an expression for the quantum Fisher information matrix
\begin{equation}
    \mathbf{H}=\begin{pmatrix}
           H_{\phi,\phi} && 0\\
           0 && H_{t,t}
    \end{pmatrix}
        ,
\end{equation}
with $H_{\phi,\phi}=4t^2(1-t^2)$, and $H_{t,t}=4/(1-t^2)$. The weak commutator, however, is non zero, thus we have to give up any hopes of finding a single measurement by which the two parameters can be estimated at once at their best possible precision.

We can now inspect our choice of a measurement, and calculate the quantities $F_{\phi,\phi}$ and $F_{t,t}$, observing a trade-off as we vary the measurement setting: for $t_m=0$ $F_{\phi,\phi}$ vanishes and $F_{t,t}=H_{t,t}$, while for $t_m=1/\sqrt{2}$, $F_{t,t}=0$, and $F_{\phi,\phi}=H_{\phi,\phi}$ for $\phi=\pi/2$. It would be wrong to think that by properly setting $t_m$, we may find a trade-off condition that allows to estimate $\phi$ and $t$: the whole Fisher information matrix must be calculated and inverted, and this reveals it is singular for any value of $t_m$. This means, through the matrix QCRB \eqref{eq:MQCRB}, that the covariance matrix diverges. No information on the individual parameters can be inferred. This is a consequence of the fact we can not resolve two parameters from only two normalised detection probabilities.

As a viable strategy, we can alternate between performing a measurement at $t_m=0$, and one at $t_m=1/\sqrt{2}$ with weight $w$ and $1-w$, respectively; this corresponds to a four-outcome generalised measurement. This eventually leads to the bounds:
\begin{equation}
    \begin{aligned}
    F^{(\rm eff)}_{\phi,\phi}= w H_{\phi,\phi},\\
        F^{(\rm eff)}_{t,t}= (1-w) H_{t,t}.
    \end{aligned}
\end{equation}
 As for the information extraction efficiency, $\Upsilon(\phi,t)$ takes the value 1, independently of $w$, to be compared with the maximum possible value 2. This means that such strategies are unable to extract all the information available in principle on the two parameters.

\subsection{Further bounds}
There is another point to be considered: we have introduced a derivative operator for quantum states in its symmetric form \eqref{SLD}, but this option is not unique. In fact, we may introduce a right logarithmic derivative (RLD) $R_h$ associated to $\phi_h$ following Yuen and Lax~\cite{1055103}
\begin{equation}
\label{RLD}
    \frac{\partial \rho_{\vec \phi}}{\partial \phi_h} = \rho_{\vec \phi}  R_h, 
\end{equation}
for a single parameter $\phi$, and an alternative Fisher information matrix as
\begin{equation}
   J_{h,k}(\vec \phi)=\text{Tr}\left[R_h^\dag \rho_{\vec \phi}R_k  \right].
\end{equation}
Notice that ${\bf J}(\vec \phi)$ is not necessarily real, and it sets the lower bound 
\begin{equation}
\label{eq:conrld}
    \Gamma\geq\frac{1}{M}\left(\text{Tr}[{\bf W}\,\text{Re}[{\bf J}^{-1}]]+\|\sqrt{{\bf W}}\,\text{Im}[{\bf J}^{-1}]\sqrt{{\bf W}} \|_1\right),
\end{equation}
where $\|{\bf A}||_1=\text{Tr}[\sqrt{{\bf A}^\dag{\bf A}}]$~\cite{Nagaoka}. In the single parameter case, it can be verified that the SLD always leads to the tightest bound~\cite{1055173}; this property does not extend to the multiparameter case, thus both cases must be inspected to assess the stricter bound. 

As an example, we can study the case of the simultaneous estimation of the real and imaginary parts of the amplitude $\alpha$ of a coherent state. We can adopt a pure state model by means of the displacement operator $\vert\alpha\rangle = e^{-\sqrt{2}i(\alpha_r\hat p -\alpha_i\hat x)}\vert 0\rangle$, where we defined $\alpha_r=\text{Re}[\alpha]$ and $\alpha_i=\text{Im}[\alpha]$~\cite{Loudon:105699,Bachor}. The generators of the two parameters are thus $G_r = \sqrt{2}\hat p$ and
 $G_i = -\sqrt{2}\hat x$, leading to a diagonal SLD Fisher information matrix:
 \begin{equation}
         \mathbf{H}=\begin{pmatrix}
           4 && 0\\
           0 && 4
    \end{pmatrix}
        .
     \end{equation}
     
RLD operators can not be constructed for pure states, however, we can make use of a theorem by Fujiwara~\cite{Fujiwara, Kok} stating that, if the SLD operators satisfy the D-invariance condition~\footnote{D-invariance is satisfied when the linear span of the SLD operators is mapped onto itself by the application $\mathcal{D}$, defined by means of $\rho_{\vec \phi}X-X\rho_{\vec \phi}=i(\rho_{\vec \phi}\mathcal{D}(X)+\mathcal{D}(X)\rho_{\vec \phi})$ for any operator $X$ in the span.}, we can nevertheless associate an RLD quantum Fisher information matrix to $\vert \psi_{\vec \phi}\rangle$ satisfying
\begin{equation}
    \mathbf{J}^{-1}=\mathbf{H}^{-1}+\mathbf{H}^{-1}\cdot\mathbf{D}\cdot\mathbf{H}^{-1},
\end{equation}
where the matrix $\mathbf D$ is the one of the weak commutators \eqref{eq:commSLD}, and for this example writes
\begin{equation}
    \mathbf{D}=
    \begin{pmatrix}
           0 && -4i\\
           4i && 0
    \end{pmatrix},
    \end{equation}
eventually leading to the expression
\begin{equation}
\mathbf{J}^{-1}=\frac{1}{4}\begin{pmatrix}
           1 && -i\\
           i && 1
    \end{pmatrix}.
\end{equation}
For equal weights, $\mathbf{W}=\mathbf{I}$, the contribution of the imaginary part of $\mathbf{J}$ in \eqref{eq:conrld} is  $\|\text{Im}[{\bf J}^{-1}] \|_1=1/2$, thus revealing how the RLD bound becomes the most informative one; this was originally highlighted by Yuen and Lax in~\cite{1055103}.

Further generalisation can be obtained by a construction due to Holevo~\cite{Holevo:1414149}, that considers families of operators $\vec X=\{X_h\}$ such that $\text{Tr}\left[X_h\frac{\partial \rho_{\vec \phi}}{\partial \phi_k}\right]=\delta_{h,k}$. Defining $Z_{h,k}[\vec X]={\text Tr}[\rho_{\vec \phi}X_hX_k]$, a lower bound is set as
\begin{equation}
\label{eq:holevo}
    \Gamma\geq\frac{1}{M}\min_{\vec X}\left(\text{Tr}[{\bf W}\text{Re}[{\bf Z}[\vec X]]]+\|\sqrt{{\bf W}}\text{Im}[{\bf Z}[\vec X]]]\sqrt{{\bf W}} \|_1\right),
\end{equation}
which is tighter than the ones obtained from the logarithmic derivatives \eqref{eq:consld} and \eqref{eq:conrld}, and can be saturated in principle, but allowing for collective measurements~\cite{PhysRevA.73.052108,Hayashi,Gill,Yang:2019qv}. Remarkably, the same D-invariance conditions ensuring the optimality of the RLD bound also grant that this corresponds to the Holevo bound~\cite{doi:10.1063/1.4945086}. Although an explicit analytical calculation of this limit is often unattainable, it can be put in the form of a semi-definite problem that allows to find numerical solutions efficiently~\cite{PhysRevLett.123.200503}.

\subsection{Applications}
Investigations of multiparameter bounds have been inspired by the possibility of accessing complex signals in communications~\cite{1055103,1055173}. In the optical domain, this problem can be recast as the estimation of a displacement in phase space, comprising a real and an imaginary part, we have discussed in the previous paragraph. This was initially studied in \cite{PhysRevA.87.012107}, in which the use of two-mode squeezed states is demonstrated to be beneficial; the experimental realisation followed shortly, highlighting the connections of this problem to that of continuous-variable dense coding~\cite{Steinlechner:2013qy}. The tightness of the SLD- and the RLD-based bounds depends on the value of the squeezing parameter $r$, with the first bound being more relevant for high squeezing; remarkably, the Holevo bound provides a unifying view of the optimal precision~\cite{PhysRevA.97.012106}.

Extending the simple two-mode interferometer case, the estimation of multiple phases $\{\phi_h\}$ in multi-arm arrangement represent a relevant application of the multiparameter approach, also because it models phase imaging of transparent objects. Since these refer to independent modes, all corresponding generators $\hat G_h$ commute, thus we may expect that these parameters can be estimated jointly at their ultimate precision: explicit calculations on fixed-photon number states~\cite{PhysRevLett.111.070403}, as well as on Gaussian states~\cite{PhysRevA.94.042342} have demonstrated this is the case. The key aspect is finding a measurement which can, in principle, saturate the matrix CRB~\eqref{eq:consld}, while delivering the values of each individual $\phi_i$. The expression for such a measurement can be found explicitly, however, finding a realistic implementation for it in the laboratory is an entirely different matter. These measurements, in fact, need obeying symmetry conditions~\cite{PhysRevLett.119.130504}, which may not be satisfied by most experimentally viable interferometers~\cite{Spagnolo12,Ciampini16}. The unavailability of such optimal schemes, however, only makes the strategies partially sub-optimal, without compromising the quantum enhancement: this has been demonstrated in the experiment in~\cite{Polino:19,polinogianani}, addressing the estimation of two phases in a three-arm integrated interferometer. The general theory for the sensitivity in such systems has been derived in~\cite{PhysRevLett.121.130503}: understanding the limits demands complementary particle and mode descriptions of the problem in order to define what are the SNL and the HL in multiphase problems, and how to attain states able to achieve those.

The many complications behind multiparameter estimation advice to restrict its use when necessary. For instance, in the example of the MZI above, it would not be worth investing resources in estimating $t$, when this parameter can be accessed with a calibration. The same applies for the loss $\eta$ of the system, including the efficiencies of the detectors. For such cases, an off-line procedure that assesses all relevant nuisances is a much more appealing option. In particular, studies have been devoted to understanding optimal estimation of loss; these have demonstrated that this task remains essentially classical, in that the scaling of the uncertainty with the resources always follows the SNL $\Delta^2\eta \sim 1/N$~\cite{PhysRevLett.98.160401,PhysRevA.79.040305}.  

When the sample is inserted, however, further loss may occur, thus the transmission of this object may seem to constitute a valid parameter to be estimated, jointly with the phase. The explicit calculations demonstrate a similar trade-off in the available Fisher information for these two parameters~\cite{PhysRevLett.123.200503}. Whether this approach is worth pursuing is dictated by the details of the problem: if loss, in our example, or any other parameter in general, may be subject to variations as phase is measured, then a multiparameter approach guarantees that the working conditions are assessed properly as the measurement evolves. This may be the case when monitoring of time-dependent parameters~\cite{Yonezawa1514}, or when spurious effects come into play~\cite{PhysRevLett.124.140501,Suzuki_2020}.

The problem of estimating small separations in images serves as an epitomising example. Lord Rayleigh put forward some simple considerations on the resolution of imaging systems~\cite{Rayleigh}, which can be summarised, in modern terms, as the impossibility to tell two point sources apart, if their point spread functions significantly overlap. In the language of parameter estimation, the Fisher information associated to the separation $d$ vanishes as $d$ approaches zero; this is commonly referred to as Rayleigh's curse~\cite{PhysRevX.6.031033}. But we have just learned that the Fisher information for one specific measurement scheme being zero does not amount to possessing no information on that parameter under all circumstances: we may have been very clumsy in choosing our measurement and, in fact, we have. Direct intensity detection is indeed a poor option, and there exist alternative schemes, notably based on coherent detection~\cite{Tsang_2017}, for which the Fisher information does not vanish. These have been first proposed in~\cite{PhysRevX.6.031033}, and then demonstrated with spatial degrees of freedom~\cite{Yang:16,PhysRevLett.118.070801,
PhysRevLett.121.250503,Zhou:19}, as well as with spectral and temporal properties~\cite{PhysRevLett.121.090501}.

In real scenarios, however, the position of the centroid of the two sources is needed, since the optimal measurement for the separation $d$ requires to know its value. Further, the protocol can be made robust against differences in the intensities from the two sources, but this unbalance should be known~\cite{Bonsma_Fisher_2019}.  The proper approach to follow is then the multiparameter scenario~\cite{PhysRevA.96.062107}. Remarkably, the available Fisher information can remain finite, even when accounting for the correlations between the parameters, as demonstrated in an experiment addressing the frequency-time domain~\cite{PRXQuantum.2.010301}. This also extends to considering the simultaneous estimation of axial and transverse separations of the two sources~\cite{PhysRevLett.121.180504,PhysRevLett.122.140505,PRXQuantum.2.020308}.

\section{Concluding remarks}

Attempting predictions about such a swiftly changing field as quantum metrology is at risk of making a spectacle of ourselves. Nevertheless, we can venture to take interpolations of current trends in order to try and ground our speculations.

The production of quantum states of light, especially squeezed states, demands a certain familiarity with nonlinear optical effects. It was thus natural to start asking questions on the potential of nonlinear evolution for quantum metrology~\cite{PhysRevA.33.4033}. Replacing beam splitters with active elements has demonstrated to deliver phase estimation beyond the SNL~\cite{Anderson:17,PhysRevLett.124.173602}, with distinctive advantages in terms of loss-tolerance~\cite{PhysRevLett.119.223604} and spatial properties~\cite{Frascella:19}.

Multiparameter scenarios can be extended to cases with a continuum of parameters, notably that of waveforms~\cite{PhysRevLett.106.090401,PhysRevX.5.031018} and, in general, varying signals in time~\cite{PhysRevLett.120.040503}. Reaching the continuous limit imposes non trivial constraints to the optimal use of resources, which must take into account uncertainties associated to both direct measurements, as well as interpolation; in turn, fundamental limits to estimation depend on the regularity of the signals~\cite{PhysRevLett.124.010507}.    

The systems to be accessed by quantum probes needs not being localised in one spatial location: extending this framework to include the monitoring of distributed systems has lead to an intense activity~\cite{PhysRevA.97.032329,PhysRevA.97.032329,PhysRevLett.120.080501,PhysRevA.97.042337,PhysRevLett.121.043604}, which is also fostering considerations on security~\cite{PhysRevA.99.022314,shettell2021cryptographic}. 

These three examples point to the direction of the dialogue of quantum metrology with different disciplines, {\it viz.} nonlinear optics, signal processing, and quantum communications. As no discipline is an island, maintaining the vitality of quantum metrology in the next future is knit to keep listening to problems and challenges coming from other fields.  

The literature of reviews on quantum metrology is rich, and can satisfy tastes and needs of all sorts. The short reviews in~\cite{GLMreviewScience,Giovannetti:2011yq} are excellent primers on concepts and methods, while \cite{ParisReview} presents more advanced material, equally precious to theorists and experimentalists to delve into the subject. Those feeling the need of more introductory material on quantum optics may turn to \cite{DEMKOWICZDOBRZANSKI2015345,doi:https://doi.org/10.1002/9781119009719.ch5}. The work in~\cite{T_th_2014} assumes some solid background in quantum information, but illustrates meticulously the connections to informational aspects.   As for multiparameter estimation, \cite{doi:10.1080/23746149.2016.1230476} offers a gentle introduction to the topic, while \cite{ALBARELLI2020126311,Kok} provide a wider overview. Focussing on photonics, the works in~\cite{Pirandola:2018qd} and especially~\cite{doi:10.1116/5.0007577} provide a comprehensive discussion on recent progress; the reviews in~\cite{RevModPhys.89.035002,RevModPhys.90.035005}, instead, are dedicated to different architecture, but present very accessible general discussions.

\section*{Acknowledgements}
 I am indebted to Animesh Datta and Marco Genoni for  countless conversations on quantum metrology that constitute the basis of this tutorial. I also acknowledge exchange with F. Albarelli, V. Cavina, A. De Pasquale, V. Giovannetti, P. Humphreys, Z. Huang, W.S. Kolthammer, C. Macchiavello, L. Maccone, M. Paris, L. Pezz\`e, E. Polino, S. D'Aurelio, L.L. S\'anchez-Soto, F. Sciarrino, A. Smerzi, N. Spagnolo, M. Vidirghin, I. Walmsley, and especially all current and past members of the NEQO group at Roma Tre: E. Roccia, V. Cimini, I. Gianani, L. Mancino, and M. Sbroscia.

This work has been supported by the European Commission via the FET-OPEN-RIA project STORMYTUNE (grant agreement number 899587).

\appendix
\section{Basic quantum optics}

We provide a short guide to the quantum treatment of light to those readers who are not entirely confident with this topic, yet have an interest in quantum metrology. This has the purpose of helping them through the calculations we have employed in the main text.

The simplest quantum treatment of light consists in introducing photons as massless particles, with degrees of freedom taken from the corresponding mode of the electromagnetic field. The photon therefore has momentum  $k=hc/\lambda$, energy $\hbar \omega$, spin $1$ related to its polarisation (with the component at $m_s=0$ being suppressed due to the transverse nature of the waves);  this description can be extended to include wavepackets in time or states with orbital angular momentum. This is akin to forcing a first quantisation treatment to the photons, which has much to left to be desired, nevertheless provides useful guidance for an intuitive understanding of experiments.

The proper way of treating the field relies on second quantisation. A harmonic oscillator is associated to each mode of the field, thus the respective Hamiltonian is $\hbar\omega\left(\hat n +1/2\right)$, where the number operator $\hat n$ has a discrete spectrum $n=0,1,2,...$, and counts the number of elementary excitations, {\it i.e.} what we have introduced as photons. Eigenstates of the energy are the number states, or Fock states, $\vert n \rangle.$ Notice how the vacuum state $\vert 0 \rangle$ is associated to a non-vanishing energy; this is responsible for observable phenomena, foremost the presence of spontaneous emission.

The number operator is written as the product of two non-Hermitean operators $\hat n = \hat a^\dag\hat a$, whose actions on number states are $\hat a^\dag \vert n\rangle = \sqrt{n+1}\vert n+1 \rangle$, and $\hat a \vert n\rangle = \sqrt{n}\vert n-1 \rangle$. Since $\hat a^\dag$ adds a photon on that mode, it takes the name of creation operator, and, for the opposite reason, $\hat a$ is called destruction or annihilation operator.
They satisfy the commutation relations $[\hat a,\hat a^\dag]=1.$ Creation and destruction operators help defining the quadrature operators $\hat x =\sqrt{N_0}\left(\hat a^\dag +\hat a\right)$, and $\hat p =i\sqrt{N_0}\left(\hat a^\dag -\hat a\right)$, which can be interpreted, as mentioned, as in-phase and in-quadrature components of the electric field with respect to a local oscillator, borrowing this picture from signal processing. Their commutation relation is then $[\hat x, \hat p]=2iN_0$. Creation and destruction operators referring to non-overlapping modes, instead, commute, and, consequently, so do quadratures.

Fock states have vanishing expectation values for the field $\langle n\vert\hat x\vert n\rangle=0$, $\langle n\vert\hat p\vert n\rangle=0$, therefore classical eletromagnetism can not be recovered in the simple limit of large-$n$ states, as no phase can be associated to the Fock states. The classical conditions of a field with given amplitude and phase are approximated by minimal uncertainty states, in the form of coherent states
\begin{equation}
\label{coherent}
    \vert \alpha \rangle = e^{-|\alpha|^2/2}\sum_{n=0}^\infty \frac{\alpha}{\sqrt n!}\vert n\rangle,
\end{equation}
where $\alpha$ is a complex number, associated to the field amplitude. Coherent states are eigenstates of the destruction operator $\hat a \vert\alpha \rangle =\alpha \vert \alpha \rangle$, and, while they are not orthogonal $\langle \beta\vert \alpha \rangle = \exp[-(|\alpha|^2+|\beta|^2-\beta^*\alpha)/2]$, they still form an overcomplete basis, in that any state $\rho$ can be written in the form
\begin{equation}
    \rho = \frac{1}{\pi}\int d^2\alpha P(\alpha)\vert \alpha \rangle \langle \alpha \vert.
\end{equation}
The function $P(\alpha)$ is known as the Glauber-Sudarshan distrubution, and can show singular behaviours, which are signature of nonclassicality. For coherent states, we get $\langle \alpha\vert\hat x\vert \alpha\rangle=2\sqrt{N_0}\text{ Re}[\alpha]$, and $\langle \alpha\vert\hat p\vert \alpha\rangle=2\sqrt{N_0}\text{Im}[\alpha]$, as we could expect from the physical meaning we attached to $\alpha$. Coherent states are found by the application of the displacement operator $\hat D(\alpha)=e^{\alpha\hat a^\dag-\alpha^* \hat a}$ to the vacuum state. The action of $\hat D(\alpha)$ is a rigid translation of the $(x,p)$ phase space so that the origin moves to the point with coordinates $x_0=2\sqrt{N_0}\text{Re}\alpha$, and  $p_0=2\sqrt{N_0}\text{Im}\alpha$, hence its name; in terms of the quadrature operators, the displacement operator is written as $\hat D(x_0+ip_0)=e^{-\frac{i}{2N_0}(x_0\hat p -p_0 \hat x)}$.

In the analysis of metrological protocols, we are often demanded to evaluate the variance of number or quadrature observables in these states. For Fock states, $n$ is clearly a well-defined number, while the quadratures have a variance $ \Delta^2x=\Delta^2p=N_0(2n+1)$. For coherent states, we have $ \Delta^2x=\Delta^2p=N_0$, regardless the amplitude $\alpha$: in the limit of large $\vert\alpha\vert$, we recover classical fields with well-defined amplitude and phase. As for the number observable in coherent states, the expression \eqref{coherent} implies a Poisson distribution with mean $\langle \alpha \vert \hat n \vert \alpha \rangle = \vert \alpha \vert^2$.

In the description of the state evolution, a rotating frame is often used not to take into account fast phase oscillations as $e^{-i\omega t}$, and the Heisenberg picture offers a more practical approach. Linear systems will induce transformations of the kind $U\hat a_iU^\dag=\sum_j c_{i,j}\hat a_j$, linking the mode $i$ with the other modes involved in the evolution. For our purposes, we mostly need to describe only two elements: the phase shifter and the beam splitter. 

An object imparting a phase shift $\phi$ implements the transformation of $\hat a$ to $e^{-i\phi}\hat a$. Consequently, the output quadratures $\hat x'$ and $\hat p'$ are rotated as
\begin{equation}
\label{phaseshift}
\begin{aligned}
    \hat x' = \cos\phi \hat x + \sin\phi \hat p\\
    \hat p' = \cos\phi \hat p - \sin\phi \hat x
    \end{aligned}
\end{equation}
A phase shift thus acts on the Fock state $\vert n\rangle$ by transforming it to $e^{-i n\phi}\vert n\rangle$, while this same operation brings the coherent state $\vert \alpha \rangle$ to $\vert e^{-i\phi}\alpha \rangle$.

A lossless beam splitter is characterised by its transmissivity $t$ and its reflectivity $r$, satisfying $|t|^2+|r|^2=1$; the relation between the phases of $r$ and $t$ depends on the chosen convention, compatibly with the unitarity of the transformation. Calling $\hat x_1$ and $\hat x_2$ the quadratures of the two input modes, these evolve to the output modes $\hat x'_1$ and $\hat x'_2$ under the action of the beam splitter as  
\begin{equation}
\label{beamsplitter}
\begin{aligned}
    \hat x'_1 &=& t \hat x_1 + r \hat x_2,\\ 
    \hat x'_2 &=& t \hat x_2 - r \hat x_1.\\ 
    \end{aligned}
\end{equation}
This implies that, given two input states in input arms, the average value of the output quadratures will be the linear combination of the input ones. As for their variances, we obtain
\begin{equation}
    \Delta^2x'_1 =t^2 \Delta^2x_1 + r^2 \Delta^2x_2+rt(\langle \hat x_1 \hat x_2 \rangle - \langle \hat x_1\rangle\langle \hat x_2\rangle),
\end{equation}
and a similar expression for $\Delta^2 x'_2$.

As an example, we consider a symmetric Mach-Zehnder interferometer imparting a relative phase shift $\phi$ between its two arms, divided up as $\phi/2$ on one mode and $-\phi/2$ on the other. It can be shown, by combining \eqref{phaseshift} and \eqref{beamsplitter}, that the action of the whole interferometer is a single beam splitter with transmittivity $t=\cos(\phi/2)$. We take a coherent state $\vert\alpha\rangle$ and a squeezed state $\vert \alpha \rangle$ as inputs; for simplicity we take $\alpha$ real, and $\vartheta=0$ in the squeezed state, corresponding to squeezing in the $\hat x$ direction. A measurement of the $\hat x'_2$ quadrature leads to an average value $\langle \hat x'_2 \rangle = 2\sqrt{N_0}\alpha \sin(\phi/2)$, with a variance $\Delta^2 x'_2=N_0\left(\cos^2(\phi/2)\, e^{-2s}+\sin^2(\phi/2) \right)$. The minimal uncertainty $\sigma^2$ on $\phi$ is found around $\phi=0$ by error propagation: 
\begin{equation}
    \left.\sigma^2=\frac{\Delta^2x'_2}{d\langle x'_2\rangle/d\phi}\right\vert_{\phi=0}=\frac{e^{-2s}}{\alpha^2},
\end{equation}
as we had found in the main text.

Besides their relevance as actual elements in the apparatus, beam splitters are also employed as effective description for loss; in this case, the second input arm is generally taken to be in the vacuum state, but when used to describe an inefficient detector, a thermal state may also model dark counts. In our example above, the efficiency $\eta$ of the detector lowers the average to $2\sqrt{\eta N_0}\alpha$, and raises the squeezed variance to $N_0(\eta e^{-2s}+1-\eta)$. Consequently, the uncertainty $\sigma^2$ is now evaluated as
\begin{equation}
    \sigma^2 = \frac{1}{\alpha^2}\left(e^{-2s}+\frac{1-\eta}{\eta}\right).
\end{equation}
Notice how the term $(1-\eta)/\eta$ appears as added noise in the fluctuations of the quadrature~\cite{QND}.

\section{Derivation of the classical Cram\'er-Rao bound}

We detail here a proof of the scalar Cram\'er-Rao bound. Two regularity conditions must hold: first $\partial \log p(x|\phi)/\partial \phi$ must be regular for all $x$ and $\phi$, second for all estimators $\tilde \phi(x)$ integration over $x$ and differentiation by $\phi$ should commute in the expression
\begin{equation}
\label{eq:commutano}
    \int dx\,\tilde \phi(x) \frac{\partial}{\partial\phi} p(x|\phi) =\frac{\partial}{\partial\phi} \int dx\,\tilde \phi(x)  p(x|\phi).
\end{equation}
We recognise that the right-hand side is the derivative of the expectation value of $\tilde \phi(x)$, {\it i.e.} $\partial(\phi+b(\phi))/\partial\phi =1+b'(\phi).$ The left-hand side of \eqref{eq:commutano} is the expectation value $\mathbf{E}[\tilde \phi(x)V(x,\phi)]$. Since the expectation value of the score is zero, the latter is also the covariance of the two statistical variables $V(x,\phi)$ and $\tilde\phi(x)$. The Cauchy-Schwartz inequality then states that the variances $\sigma^2=\mathbf{V}[\tilde \phi (x)]$ and $F(\phi)=\mathbf{V}[V (x,\phi)]$ are bounded from below by their covariance 
\begin{equation}
    \mathbf{V}[\tilde \phi(x)]\mathbf{V}[V(x,\phi)]\geq \mathbf{E}[\tilde \phi(x)V(x,\phi)],
\end{equation}
implying through Eq.~\eqref{eq:commutano} 
\begin{equation}
\label{eq:protoCRB}
\sigma^2F(\theta)\geq (1+b'(\phi))^2. \end{equation}
Since the Fisher information is additive for independent events, we have that, after $M$ repetitions, the total information is $MF(\theta)$. This leads to the expression
\begin{equation}
    \sigma^2 \geq \frac{(1+b'(\phi))^2}{M F(\phi)},
\end{equation}
which reduces to the usual bound \eqref{ccrb} for unbiased estimators.

\section{Derivation of the quantum Cram\'er-Rao bound}

The expression for the scalar quantum Cram\'er-Rao bound is found as follows. For quantum states, the probability distributions $p(x|\phi)$ are obtained by Born's rule: $p(x|\phi)=\text{Tr}[\rho_\phi \Pi_x]$, with $\Pi_x$ the measurement operator associated to the outcome $x$. By the definition of the SLD \eqref{SLD}, we find that
\begin{equation}
\frac{\partial p(x|\phi)}{\partial \phi} =\text{Re}\left[ \text{Tr}[\rho_\phi L_\phi\Pi_x]\right].
\end{equation}
Any strategy has then a Fisher information which is limited above as
\begin{equation}
\label{eq:meserve}
\begin{aligned}
    F(\phi)&\leq \int dx\,\frac{1}{\text{Tr}[\rho_\phi\Pi_x]} \left\vert \text{Tr}[\rho_\phi L_\phi\Pi_x] \right\vert^2\\
    &= \int dx\,\frac{\left\vert\text{Tr}[\sqrt{\Pi_x}\sqrt{\rho_\phi}\sqrt{\rho_\phi}L_\phi \sqrt{\Pi_x}]  \right\vert^2} {\text{Tr}[\rho_\phi\Pi_x]}.
    \end{aligned}
\end{equation}
The last step is needed to employ the Cauchy-Schwartz inequality to the scalar product of matrices $\vert\text{Tr}[X^\dag Y]\vert^2\leq \text{Tr}[X^\dag X]\text{Tr}[Y^\dag Y]$. Taking ${X=\sqrt{\Pi_x}\sqrt{\rho_\phi}}$ and ${Y=\sqrt{\rho_\phi}L_\phi\sqrt{\Pi_x}}$, we get
\begin{equation}
\begin{aligned}
    F(\phi)&\leq \int dx\,\text{Tr}[\rho_\phi  L_\phi \Pi_x L_\phi ]\\
    &=\text{Tr}[ L_\phi^2 \rho_\phi], \end{aligned}
\end{equation}
where we have used the fact that the measurement operators form a resolution of the identity. This can be employed to show the bound \eqref{qcrb}. The optimality of the eigenbase of $L_\phi$ as the measurement choice can be demonstrated by direct substitution, after observing that $\text{Tr}[\rho_\phi L_\phi \Pi_x]$ must be real for the inequality \eqref{eq:meserve} to be saturated: this is not restrictive, as $L_\phi$ can be taken as Hermitean. Since now $\Pi_x =\vert x\rangle\langle x\vert$, and $L_\phi\vert\phi\rangle = l_x \vert x \rangle$, we obtain
\begin{equation}
    F(\phi)=\int dx\, l_x^2 \langle x\vert\rho_\phi\vert x\rangle = \text{Tr}[L^2_\phi \rho_\phi],
\end{equation}
concluding our proof.

\bibliography{tutorial}

\end{document}